\documentclass[usenatbib]{mn2e}

\usepackage{astrojournals}
\usepackage{graphicx}
\usepackage{multirow}
\usepackage{url} 
\usepackage{amsmath}
\usepackage{times}
\usepackage{amssymb}
\usepackage{xcolor}
\def\music{\texttt{MUSIC}}

\def\gadgetF{\texttt{GADGET-2}}
\def\ahf{\texttt{AHF}}
\def\vmax{\mathrm{V}_{\rm max}}
\def\rvir{\mathrm{r}_{\mathrm{vir}}}
\def\mvir{\mathrm{M}_{\mathrm{vir}}}

\def\fbar{f_{\rm bar}}
\def\fgas{f_{\rm gas}}
\def\fstar{f_*}

\def\mobstar{\mathrm{M}_{*}}

\def\FeH{[Fe/H]}
\def\reobstar{\mathrm{r}_{1/2}^{*}}

\def\mtotTT{\mathrm{M}_{500}^{\mathrm{tot}}}
\def\mdmTT{\mathrm{M}_{500}^{\mathrm{dark}}}
\def\mbarTT{\mathrm{M}_{500}^{\mathrm{bar}}}
\def\mgasTT{\mathrm{M}_{500}^{\mathrm{gas}}}
\def\mstarTT{\mathrm{M}_{500}^{*}}

\newcommand{\msun}{{\rm M}_{\odot}}
\newcommand{\mstar}{{\rm M}_{\star}}
\newcommand{\lcdm}{$\Lambda$CDM}

\title[Forged in FIRE]{Forged in FIRE: cusps, cores, and baryons in low-mass dwarf galaxies}
\author[J.~O\~norbe et al.]{
\parbox[t]{\textwidth}{ 
Jose O{\~{n}}orbe\thanks{E-mail: \texttt{onorbe@mpia.de}}$^{1,2}$,
Michael Boylan-Kolchin$^3$, 
James S.~Bullock$^1$,
Philip F.~Hopkins$^4$,
Du\v{s}an Kere\v{s}$^5$,
Claude-Andr{\'e} Faucher-Gigu{\`e}re$^6$, 
Eliot Quataert$^7$,
Norman Murray$^8$
} 
\vspace*{6pt} \\
$^1${Department of Physics and Astronomy, University of California at Irvine, Irvine, CA 92697, USA}\\
$^2${Max-Planck-Institut fuer Astronomie, Koenigstuhl 17, 69117 Heidelberg, Germany}\\
$^3${Department of Astronomy and Joint Space-Science Institute, University of Maryland, College Park, MD 20742-2421}\\
$^4${TAPIR, Mailcode 350-17, California Institute of Technology, Pasadena, CA 91125, USA} \\
$^5${Department of Physics, Center for Astrophysics and Space Science, University of California at San Diego, 9500 Gilman Drive, La Jolla, CA 92093} \\ 
$^6${Department of Physics and Astronomy and CIERA, Northwestern University, 2145 Sheridan Road, Evanston, IL 60208, USA} \\ 
$^7${Department of Astronomy and Theoretical Astrophysics Center, University of California Berkeley, Berkeley, CA 94720} \\
$^8${Canadian Institute for Theoretical Astrophysics, 
60 St.\ George Street, University of Toronto, ON M5S 3H8, Canada} \\
\vspace{-0.5cm}
}

\begin{document}
\label{firstpage}

\maketitle

\begin{abstract}
We present multiple ultra-high resolution cosmological hydrodynamic simulations of $\mstar \simeq 10^{4-6.3} \msun$ dwarf galaxies that form within two $\mvir = 10^{9.5-10} \msun$ dark matter halo initial conditions. Our simulations rely on the FIRE implementation of star formation feedback and were run with high enough force and mass resolution to directly resolve structure on the $\sim 200$ pc scales.
The resultant galaxies sit on the $\mstar$ vs. $\mvir$ relation required to match the Local Group stellar mass function via abundance matching.  They have bursty star formation histories and also form with half-light radii and metallicities that broadly match those observed for local dwarfs at the same stellar mass.  We demonstrate that it is possible to create a large ($\sim 1$ kpc) constant-density dark matter core in a cosmological simulation of an $\mstar \simeq 10^{6.3} \msun$ dwarf galaxy within a typical $\mvir = 10^{10} \msun$ halo -- precisely the scale of interest for resolving the Too Big to Fail problem.  However, these large cores are not ubiquitous and appear to correlate closely with the star formation histories of the dwarfs: dark matter cores are largest in systems that form their stars late ($z \lesssim 2$), after the early epoch of cusp building mergers has ended.
Our $\mstar \simeq 10^{4} \msun$ dwarf retains a cuspy dark matter halo density profile that matches that of a dark-matter only run of the same system. Though ancient, most of the stars in our ultra-faint form after reionization; the UV field acts mainly to suppress fresh gas accretion, not to boil away gas that is already present in the proto-dwarf.
\end{abstract}


\section{Introduction}
\label{sec:Introduction}

Many of the most pressing problems associated with the standard LCDM paradigm  
concern the faintest $\mstar \simeq10^{6} \msun$  dwarf galaxies and
the dark matter halos that have the right abundance to host them: $\mvir \simeq 10^{10} \msun$ \citep{GarrisonKimmel:2014,Brook:2014}.
If LCDM is correct, then the dark matter halos hosting these dwarfs must be extremely inefficient
at converting baryons into stars  \citep[][]{Klypin:1999}
and they also must be significantly less dense in their centers than predicted in dissipationless 
LCDM simulations \citep[][]{BoylanKolchin:2011,MBK12,Ferrero2012,GarrisonKimmel:2014b,Tollerud:2014,Klypin2014,Papastergis:2015}. 
This latter issue (known as the Too Big to Fail problem) may be related to
indications that dwarf galaxies reside within dark matter halos that have cored density profiles 
rather than the cuspy NFW-like profiles predicted in CDM simulations \citep[][but see Strigari et al. 2014]{Flores:1994,KuziodeNaray:2008,deBlok:2008,Oh:2008,Walker:2011,Salucci:2012,Amorisco:2014,Ogiya:2015}. \nocite{Strigari:2014}

 While some authors have taken these discrepancies as motivation to explore
 non-standard dark matter models
 \citep[][]{Maccio:2010,Vogelsberger:2012,Rocha:2013,Horiuchi:2014,Governato:2015},
 others have argued that that it may be possible to naturally resolve them
 through a better understanding of star formation and feedback in low-mass
 galaxies.  Specifically, the inefficiency of dwarf galaxy formation is believed
 to be driven by supernovae feedback and the effects of an ionizing background
 \citep[][]{Dekel:1986,Bullock:2000}.  Likewise, dark matter halos may be
 transformed from cusps into cores if enough energy can be injected into the
 orbits of dark matter particles during rapid starburst events
 \citep[][]{Navarro:1996,Governato:2010,Pontzen:2012,Ogiya:2014}.  As pointed out by
 \citet{Penarrubia:2012}, these two requirements are at odds with each other:
 the need to lower the efficiency of star formation means that there will be
 less supernovae energy available to create dark matter cores.  Solving the two
 problems simultaneously therefore represents a significant theoretical
 challenge \citep{GarrisonKimmel:2013}.

Reproducing even the broad-brush properties of dwarfs in a cosmological framework, regardless of their internal structure, has been historically 
challenging. At these scales,  the relationship between stellar mass and halo mass derived from  local galaxy counts
 \citep{GarrisonKimmel:2014b,Brook:2014} implies a suppression
of galaxy formation by a factor of about $1000$.
While it is generally believed that stellar feedback is the main agent responsible
for this suppression, actually getting a physically realistic model of the relevant processes to
manifest these expectations has proven difficult.  

The past several years have proven fruitful in this regard, with many published studies achieving substantial suppression in
the conversion of baryons to stars on the scale of dwarf galaxy halos
\citep{Governato:2010,Sawala:2011,Simpson:2013,Munshi:2013,Governato:2015,TrujilloGomez:2015}.
As we show below, however, many of these studies have not quite reached the level of suppression that seems to 
be required by local galaxy counts.  Moreover, 
whether or not these feedback models also match the different observed scaling relations
for these systems \citep{Wolf:2010,Kirby:2013,Collins:2014} is still not clear. Reproducing both the correct stellar mass
and structural properties has proven to be an even more difficult challenge
\citep{Sales:2010}. The observed stellar 
metallicity - stellar mass tight correlation \citep{Gallazzi:2005,Kirby:2013} can also put very important
constraints on the feedback models and how these are implemented.

 As for the question of feedback-driven core formation, much remains debated.  Some of the most successful simulations
at producing cores in dwarf galaxies have suggested a transition mass below
 $\mstar \sim 10^7\,\msun$ where core formation becomes difficult \citep{Governato:2012}.
Using a slightly different set of simulations,  \citet{DiCintio:2014} find similar
results,  and suggest that the cusp-core transition should be most effective
when the ratio of stellar mass to dark matter halo mass relatively high, 
in massive dwarfs with $\mstar \sim 10^8\,\msun$ and $\mvir \sim 3 \times 10^{10.5}\,\msun$.
Importantly, they also find that cuspy profiles are retained for the $\mstar \simeq 10^6 \msun$ dwarfs of concern
(residing in $\mvir = 10^{10} \msun$ halos) although resolution may have been an issue in these cases. 
At some mass scale, galaxy formation may become effectively stochastic \citep[e.g., ][]{BoylanKolchin:2011}.
Recent work by \citet{Sawala:2015,Sawala:2014}, however, suggests that the scale at which stochasticity
becomes important is somewhat lower ($\mvir \sim10^{9}-10^{9.5} \msun$).

Though the results of \citet{DiCintio:2014} and \citet{Governato:2012} agree reasonably well,
a different set of high resolution simulations with a simpler
implementation of stellar feedback
have not produced cores in dwarf galaxy halos at any mass \citep{Vogelsberger:2014}, 
even though a number of other observables are well matched. 
The absence of cores produced by stellar feedback in these simulations could be due to
the fact that their sub-grid ISM and star formation model leads to star formation
histories that are (likely) artificially smoothed in time, compared to the bursty star
formation histories found in more explicit models \citep{Hopkins:2014,Muratov:2015}.
Conversely, \citet{TrujilloGomez:2015} found that radiation pressure from massive stars
was the most important source of core formation in their simulations, not thermal feedback from supernova, 
which has been the primary mode used by other groups that have produced cores.  
More generally, models for feedback that have been used up until now have been sub-grid 
and necessitated ad-hoc approximations, such as turning off cooling for material heated by SNe. 
As such it  is not clear whether the feedback we actually expect from stellar evolution models is capable
of producing large cores, or whether the mass-limit for core formation is robust.

In this paper, we attempt to minimize the freedom of sub-grid galaxy formation
models and to incorporate as many important physical processes in a manner that
is as realistic as possible at present in order to understand if and how star
formation affects the gravitational potential wells of dwarf dark matter
halos.  To these ends, we have conducted a series of high resolution cosmological 
hydrodynamical simulations of two dwarf halos using the code presented in \citet{Hopkins:2014}.
In this work, we showed that this implementation of stellar feedback successfully 
reproduces the observationally-inferred relationship between the stellar mass-dark
matter halo mass ($\mstar$-$M_{\rm halo}$) and star formation histories of 
galaxies at all redshifts where observational constraints are currently available.
\citet{FaucherGiguere:2015} recently showed that it also replicates
the neutral hydrogen content of high-redshift halos.

To our knowledge, the set of simulation presented here include the current highest resolution simulation
of this type with an explicit implementation of feedback yet achieved. 
This not only facilitates a more accurate treatment of astrophysical processes 
but is also crucial in the context of dwarfs
as dark matter probes.  The dwarfs of concern have half-light radii of $\sim 500$ pc, and thus
any dark matter core of relevance needs to be dynamically resolved at this scale.
According to well-documented convergence test studies 
\citep{Power:2003}, many previous simulations that have reported core formation on this scale were quite
 poorly resolved, some at only $\sim2-3$ softening lengths.  In what follows we make every
 effort to clarify our resolution limitations.
 
The paper is organized as follows. In Section~\ref{sec:sims}, we describe the
computational methods which we have used and our choice of initial conditions.
We present the results of our simulations in Section~\ref{sec:propz0}. We pay 
closer attention to the matter content of our simulated dwarfs, and the possible
formation of cores, in Section~\ref{sec:dmstruc}. We conclude with a summary 
where we discuss the achievements and shortcomings of the simulations in Section~\ref{sec:conc}.

\section{Simulations}
\label{sec:sims}

\begin{figure*} 
\begin{center}
\includegraphics[width=0.33\textwidth]{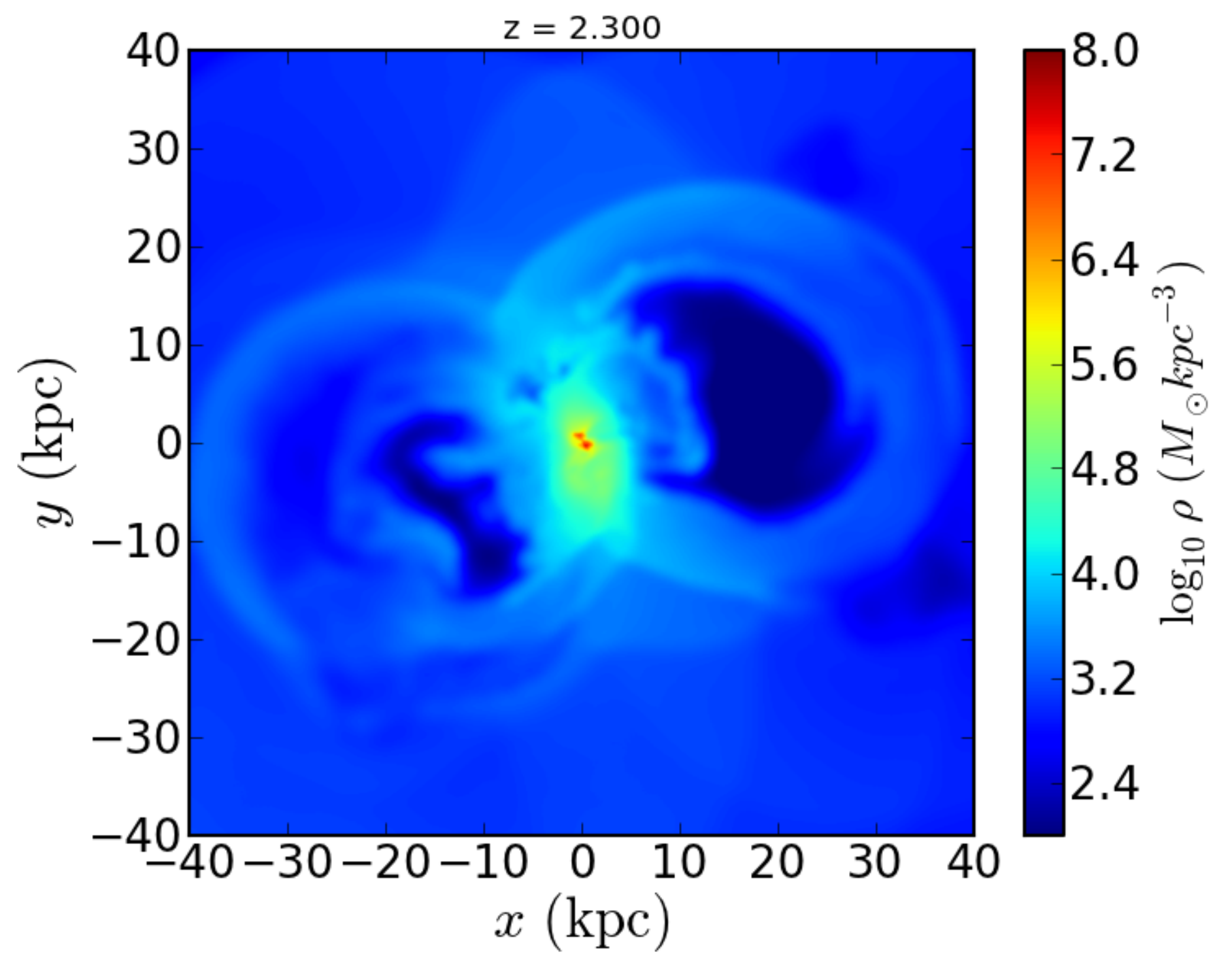}
\includegraphics[width=0.33\textwidth]{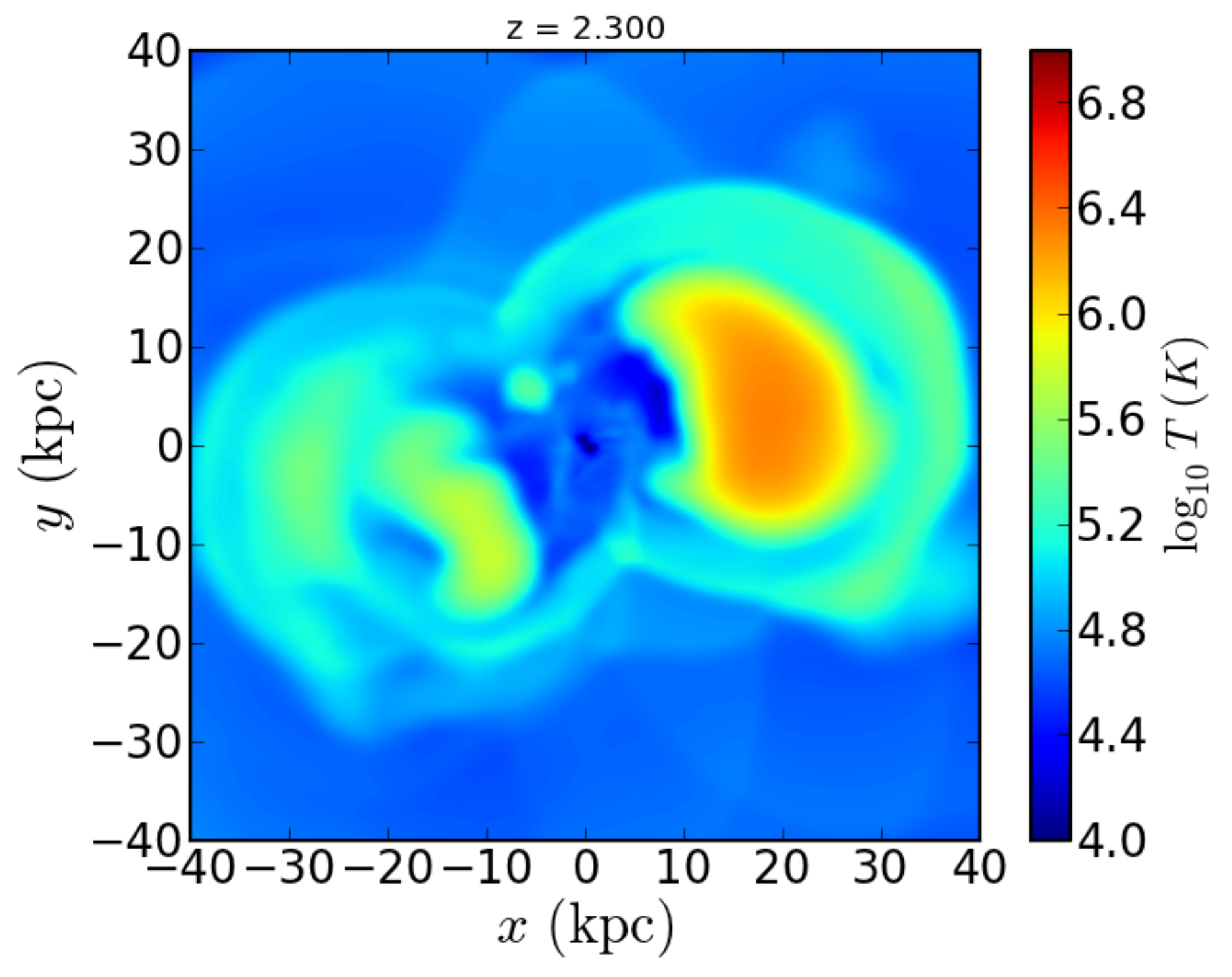}
\includegraphics[width=0.33\textwidth]{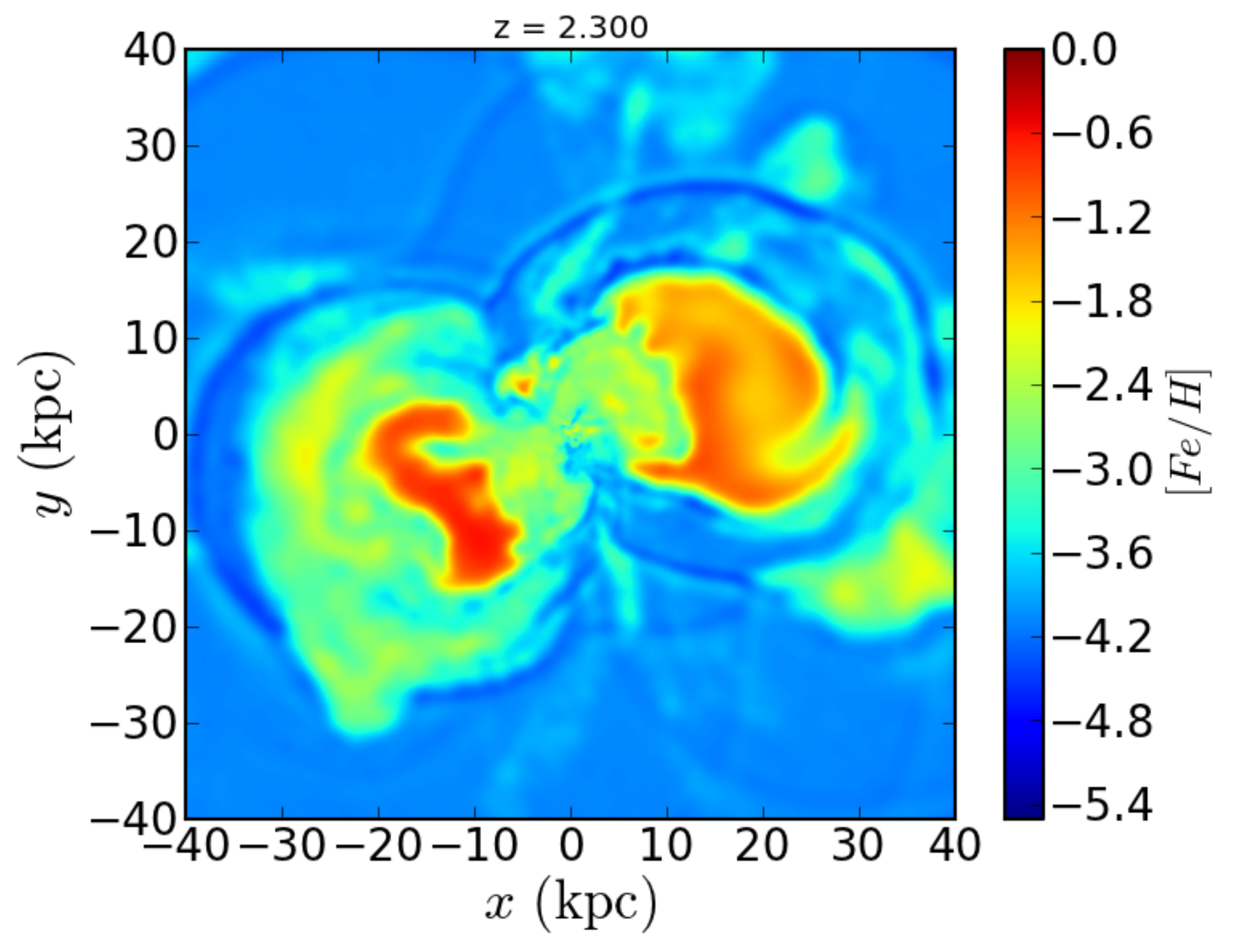}
\end{center}
\caption{From left to right, visualizations of the gas density, gas temperature and gas metallicity for the Dwarf\_early run at $z=2.3$. All panels show the same thin slice along the $z$-axis centered at the main halo. The signatures of a recent stellar burst episode are clear in all of them.}
\label{fig:vis}
\end{figure*}

We have run a series of multimass cosmological hydrodynamical simulations \citep{Porter:1985,Katz:1993} following the formation and evolution of structure in the $\Lambda$CDM model of two dwarf galaxy halos.
Each simulation is a cosmological zoom-in that includes high-resolution gas and dark matter for the flow converging region that generates the main object. The rest of the simulation box is sampled by low-resolution dark matter particles that account for tidal forces. The cosmological model adopted throughout this paper is based on cosmic microwave background results \citep{Komatsu:2011}: $\sigma_{\rm 8}=0.801$, $\Omega_{\rm \Lambda}=0.734$, $\Omega_{\rm m}=0.266$, $\Omega_{\rm b}=0.0449$, $n_{\rm s}=0.963$ and $h=0.71$.

To generate the cosmological initial conditions we made use of \music{}, an OPENMP parallel algorithm to generate multi-scale initial conditions with multiple levels of refinements for cosmological ``zoom" simulations \citep[\music{}][]{Hahn:2011} and we followed the method outlined in \citet{Onorbe:2014}. To select our dwarf candidates we first run a medium-resolution dark-matter only cosmological simulation using \gadgetF{} \citep{Springel:2005} with a cubic volume of $7$ Mpc on a side with particle mass $m_{\rm p}=9.7 \times 10^{4} \msun$ and Plummer equivalent force softening length of $176$ pc. To be able to study the main statistical properties of dwarf galaxy halos we also run a bigger dark-matter only simulation of $35$ Mpc on a side with particle mass $m_{\rm p}=1.2 \times 10^{7} \msun$ and Plummer equivalent force softening length of $563$ pc.
In this work we present simulations of two dwarf galaxy halos, one with a virial mass of $\mvir=3.2\times 10^{9} \msun$ and the other with $\mvir=9.2\times 10^{9} \msun$\footnote{Unless otherwise stated, in this paper we define the virial overdensity using the spherical top hat collapse approximation by \citet{Bryan:1998}.}. 
Based on our analysis of the $35$ Mpc simulation, we have chosen our dwarf candidates to lie as close as possible to the mean values of spin, concentration and halo formation time for its mass while still having a small Lagrangian volume \citep[see][]{Onorbe:2014}. The specific values of these parameters for our two halos can be found in Table~\ref{tab:simsinfo}. We point to Appendix~\ref{app:dmevol} for a more detailed description of these parameters and how they compare with a sample of halos in the same mass bin.

To check the convergence of our results we have run two resolution levels for our simulations: in our low-resolution hydrodynamical testing runs we use a dark matter particle mass of $1.01\times 10^{4}$ $\msun$ and a particle gas mass of $2.04\times 10^{3}$ $\msun$ (the mass resolution for the collisionless run is therefore $1.22\times 10^{4}$ $\msun$). The high resolution runs used a dark matter particle mass of $1.26\times 10^{3}$ $\msun$ and a gas particle mass of $254$ $\msun$ (the particle resolution for the collisionless run is therefore $1.5\times 10^{3}$ $\msun$).
None of the high resolution regions of the simulations presented in this work are contaminated by low resolution particles at any redshift within 1.6 virial radii.

The simulations presented in this paper use GIZMO\footnote{http://www.tapir.caltech.edu/~phopkins/Site/GIZMO} \citep{Hopkins:2014b}, run in P-SPH mode which include physical models for star formation and stellar feedback presented in \citet{Hopkins:2014}. Two of the runs presented here Ultrafaint and Dwarf\_early were also presented in \citet{Hopkins:2014} (m09 and m10 respectively).  
 We summarize their properties below, but readers interested in further details (including resolution studies and a range of tests of the specific numerical methodology) should see \citet{Hopkins:2012,Hopkins:2013b,Hopkins:2014}.

For the halo identification in the simulation we have used the public code Amiga Halo Finder \citep[\ahf{}][]{Knollmann:2009}, an MPI parallel code for finding gravitationally bound structures in simulations of cosmic structure. Results presented in this work use a highest density peak+sigma-clipping method to find the center. We have also tested different centering algorithms to confirm that our results do not depend on which method was used\footnote{A simple center-of-mass algorithm was the only method that we found not able to track the center of our systems with the accuracy required for this work.}.

\subsection{Numerical Methods}

The P-SPH method adopts the Lagrangian ``pressure-entropy" formulation of the SPH equations developed in \citep{Hopkins:2013a};
this eliminates the major differences between SPH, moving mesh, and grid
(adaptive mesh) codes, and resolves the well-known issues with
fluid mixing instabilities in previously-used forms of SPH 
\citep[e.g.][]{Agertz:2007,Sijacki:2012}. 
P-SPH also manifestly conserves momentum, energy, angular
momentum, and entropy.
The gravity solver is a heavily modified version of the GADGET-3
\citep{Springel:2005} hybrid tree-particle mesh (Tree-PM) method; but
GIZMO also includes substantial improvements in the artificial viscosity,
entropy diffusion, adaptive timestepping, smoothing kernel,
and gravitational softening algorithm, as compared to the ``previous
generation" of SPH codes. These are all described in detail in \citet{Hopkins:2014,Hopkins:2014b}.
In particular, in ``traditional" GADGET, softenings are not adaptive, and pairwise interactions are simply
smoothed by the larger of the two particle softenings. We have also modified the softening
kernel as described therein to represent the exact solution for the potential of the SPH 
smoothing kernel. Therefore our ``standard" simulations use adaptive gravitational
softening lengths for gas which minimum is a factor $\sim10$ smaller than the fixed dark matter gravitational 
softening lengths. In order to test this approach we have also run the same initial conditions
using identical softenings for both the baryonic and dark matter particles (close to the higher
dark matter default value). We labeled these runs according to the late star formation history 
of the high resolution runs (see Table~\ref{tab:simsinfo} and the discussion below for more details).


In our simulations, gas follows an ionized+atomic+molecular cooling curve from $10-10^{10}$ K, including metallicity-dependent fine-structure and molecular cooling at low temperatures, and high-temperature ($\gtrsim 10^{4}$ K) metal-line cooling followed species-by-species for 11 separately tracked species. At all times, the appropriate ionization states and cooling rates are tabulated from a compilation of CLOUDY runs, including the effect of a uniform but redshift-dependent photo-ionizing background computed in \citep{FaucherGiguere:2009}\footnote{Publicly available at http://galaxies.northwestern.edu/uvb}, together with local sources of photo-ionizing and photo-electric heating. Self-shielding is accounted for with a local Sobolev/Jeans-length approximation (integrating the local density at a given particle out to a Jeans length to determine a surface density $\Sigma$, then attenuating the background seen at that point by $\exp(\kappa_{\rm \nu}\Sigma)$).

Star formation is allowed only in dense, molecular, self-gravitating regions
above $n > n_{\rm crit}$ ($n_{\rm crit} = 100\,{\rm cm}^{-3}$ for our high-resolution simulations\footnote{The simulations get to a maximum density of $\sim10^{4}$ ${\rm cm}^{-3}$}). This threshold is much higher than that adopted in most ``zoom-in" simulations of galaxy formation (the high value allows us to capture highly clustered star formation). We follow \citet{Krumholz:2011} to calculate the molecular fraction $f_{\rm H2}$ in dense gas as a function of local column density and metallicity, and allow SF only from molecular gas. We also follow \citet{Hopkins:2013b} and restrict star formation to gas which is locally self-gravitating, i.e. has $\alpha \equiv \delta v^{2} \delta r/G m_{\rm gas}(< \delta r) < 1$ on the smallest available scale ($\delta r$ being our force softening or smoothing length). This forms stars at a rate $\dot{\rho}_{̇\star} = \rho_{\rm mol}/t_{\rm ff}$ (i.e. 100\% efficiency per free-fall time); so that the galaxy and even kpc-scale star formation efficiency is not set by hand, but regulated by feedback (typically at much lower values). 


Feedback from stellar evolution is modeled by implementing energy, momentum, mass, and metal return from radiation, supernovae, stellar winds, and photoionization. Every star particle is treated as a single stellar population, with a known age, metallicity, and mass. Then all feedback quantities (the stellar luminosity, spectral shape, SNe rates, stellar wind mechanical luminosities, metal yields, etc.) are tabulated as a function of time directly from STARBURST99 stellar population synthesis model \citep{Leitherer:1999}, assuming a \citet{Kroupa:2002} IMF. Details on the implementation of each of these physical processes in our simulations can be found in \citep{Hopkins:2014}. No black hole physics has been considered in these simulations.

Despite taking all our inputs directly from stellar population models, there are some ambiguities in how we implement them. For example, when we deposit mass, momentum, and energy to particles within the SPH kernel, we can do so according to a mass-weighting or volume-weighting scheme. We have experimented with both, and we refer to these options as {\em Feed-M} and {\em Feed-V}, respectively. 

We stress that the systematic differences due to these (and other similar) purely numerical choices \citep[see Appendix A of][]{Hopkins:2014} are relatively small for integrated quantities like the stellar mass. However, since the dynamics of galaxies and star formation are chaotic, a small perturbation can make a non-negligible difference to the shape of the star formation history. These essentially stochastic variations will provide a useful means for us to examine the role of different star formation histories in shaping cores. 

We have found that the main global parameters describing the dwarf galaxies are quite robust regardless of resolution, softening and other minor changes in the code. See Appendix~\ref{app:convergence} for a full discussion on the convergence of our results. 

\subsection{Sample Summary}

A summary of all the relevant parameters used in the ultra high resolution runs presented in this work is shown in Table~\ref{tab:simsinfo} along with the naming conventions we have adopted. In this work we present a total of six high resolution simulations of two dwarf galaxy halos, one with a virial mass of $\mvir=3.2\times 10^{9} \msun$ and the other with $\mvir=9.21\times 10^{9} \msun$ (as measured in the high-resolution collisionless simulations). For the more massive halo we present here four runs, a high resolution collisionless run (Dwarf\_dm) and a total of three different hydrodynamical runs which include two feedback implementation tests and the softening test mentioned above. 

We have named the three hydrodynamical Dwarf simulations based on their star
formation histories (see section 3.2 below).  The run we call ``Dwarf\_early''
shows most of its star formation at early times and corresponds with the
feedback method {\em Feed-V}.  The run we call ``Dwarf\_late'' uses feedback
method {\em Feed-M} and shows a more significant star formation rate at low
redshifts. The ``Dwarf\_middle'' run is the softening test which uses feedback
method ``Feed-M'' and its star formation rate history stands just between the
two. Simulations of the same dwarf using the ``Meshless Finite Mass'' method
implemented in GIZMO \citep{Hopkins:2014b} and the feedback {\em Feed-V})
method produce results very similar
to the ``Dwarf\_early'' run presented here (Fitts et al., in preparation).

For the smaller halo we have run the same number of simulations as we have for
the larger one, but their results were so similar that we present only one
hydrodynamic run (Ultrafaint, which uses {\em Feed-V}) and one collisionless run
(Ultrafaint\_dm). The hydrodynamical runs Dwarf\_early and Ultrafaint were
already presented in the first FIRE paper \citep{Hopkins:2014}.

\begin{figure} 
\begin{center}
\includegraphics[width=0.45\textwidth]{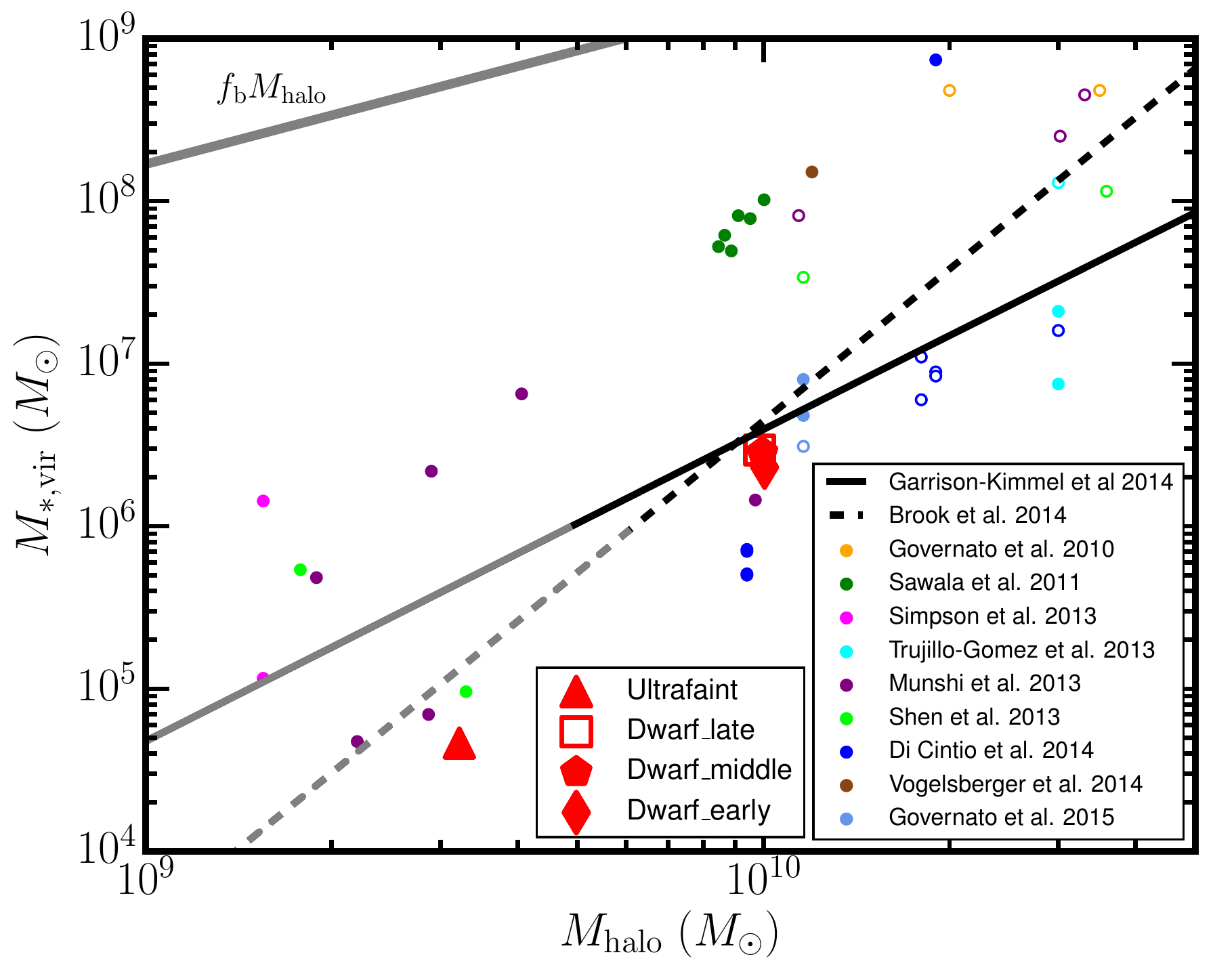}
\end{center}
\caption{Galaxy stellar mass-halo mass relation at $z = 0$ for the simulations presented in this work. Filled red triangle stands for the Ultrafaint run. Empty red square, filled red pentagon and filled red diamond stand for the Dwarf\_early, Dwarf\_middle and Dwarf\_late runs respectively. Circles stand for other published simulations in the literature using sub-grid stellar feedback models.
Empty symbols stand for runs in which a dark matter core was found according to the typical definition used in previous works: a slope $a\gtrsim-0.9$ for the dark matter density profile between 1\% - 2\% of the virial radius. The two black lines show abundance matching relations derived by Brook et al. and Garrison-Kimmel et al. using galaxy counts within the Local Volume.  These relations should be complete to $\mstar \sim 10^6 \msun$ (much deeper than other published abundance matching relations and therefore the most relevant for this comparison). To make all simulation data and abundance matching results consistent between each other, we have plotted the total virial masses of the dark matter only runs defined as $\Delta_{\rm vir}=200\rho_{\rm crit}$. A small correction has been applied whenever these values were not available in the literature (with $\leqslant10\%$ effects). In order to facilitate direct comparison with other results in the literature, the stellar mass plotted is all of the stellar mass inside the virial radius. Using a smaller radius makes only a small difference for our simulated dwarf galaxies.}
\label{fig:stellarhaloz0}
\end{figure}

As pointed out above, we have also checked the convergence of these results with
resolution by running all these setups also at a lower resolution level. We
discuss these runs in detail in Appendix~\ref{app:convergence}.  We have also
run many more ($\sim 50$) simulations at this lower resolution level of these
halos to test other purely numerical issues and the effects of adding/removing
each feedback mechanism in turn. Some of these are summarized in
\citet{Hopkins:2014}. We will not discuss them further in this paper because
they are either not instructive for the study of this work because the included
physics is not complete or because there is no change in the results. 
Even given excellent force and mass resolution, the time-step
criterion used in simulations is always a concern if many, many orbits of N-body
particles must be followed (as in the halo centers of the systems studied
here). These can artificially deteriorate a central cusp, if an insufficiently
stringent timestep criterion and/or error tolerance for the long-range force
computations is used. We have therefore re-run a subset of our low-resolution
runs, making the timestep criterion a factor of $\sim30$, and force error
tolerance a factor of $\sim100$ times more strict than our default choices. This
amounts to taking $<100$ year timesteps, with a tree force accuracy a factor
$\sim1000$ stricter than used in \citet{Governato:2012}, and a factor $\sim100$
stricter than was found to give good convergence in idealized comparisons of
dark matter zoom-in simulations in \citep{Kim:2014}. Given our very strict
default tolerances, this gave well-converged results.

Figure~\ref{fig:vis} shows visualizations of the gas density (left panel), gas temperature (middle panel) and gas metallicity (right panel) for the Dwarf\_early run at $z=2.3$. All panels show the same thin slice along the $z$-axis centered at the main halo. The signatures of a recent SN episode are clear in all of them.

\section{Results}
\label{sec:propz0}

\subsection{Basic Properties at $z=0$} 
Table~\ref{tab:simsinfo} presents some relevant parameters describing the properties of each simulation presented in this work that will allow an immediate comparison with previous simulations and observations of dwarf galaxies. 

Of particular interest is the resultant stellar mass in each dwarf.   Figure~\ref{fig:stellarhaloz0} presents the stellar mass - halo mass relation for the four hydrodynamical runs described above (large red points) compared to
the most recent estimates for this relation from abundance-matching exercises in the Local Group \citep[][black solid and dashed lines respectively]{GarrisonKimmel:2014,Brook:2014}.  The known sources of stellar feedback we include, with no adjustment, automatically produces galaxy stellar masses that are consistent with those required to match local galaxy counts\footnote{Abundance matching results below $\sim 10^{6} \msun$ stellar mass are extrapolation as observations throughout the Local Group are not complete below this limit}. For the halo mass range presented here, this is particularly impressive, as the integrated stellar mass is suppressed by factors of $\sim 1000$ relative to the Universal baryon fraction (upper solid gray line).

The smaller points in Figure~\ref{fig:stellarhaloz0} show results from previous hydrodynamical simulations of dwarf galaxies \citep{Governato:2010,Sawala:2011,Simpson:2013,Munshi:2013,Shen:2014,TrujilloGomez:2015}. The open points are those that have reported at least mild flattening of the central dark matter cusp in response to feedback effects.  We note that all of those open points are associated with systems that have formed a fair number of stars, with $M_{*} \gtrsim 7 \times 10^6$ M$_\odot$ -- more massive than the systems of concern for the Too Big to Fail Problem.  As we discuss below, one of our runs (Dwarf\_late, open square) produces a large core while forming significantly fewer stars.

\begin{figure*} 
\begin{center}
\includegraphics[width=0.45\textwidth]{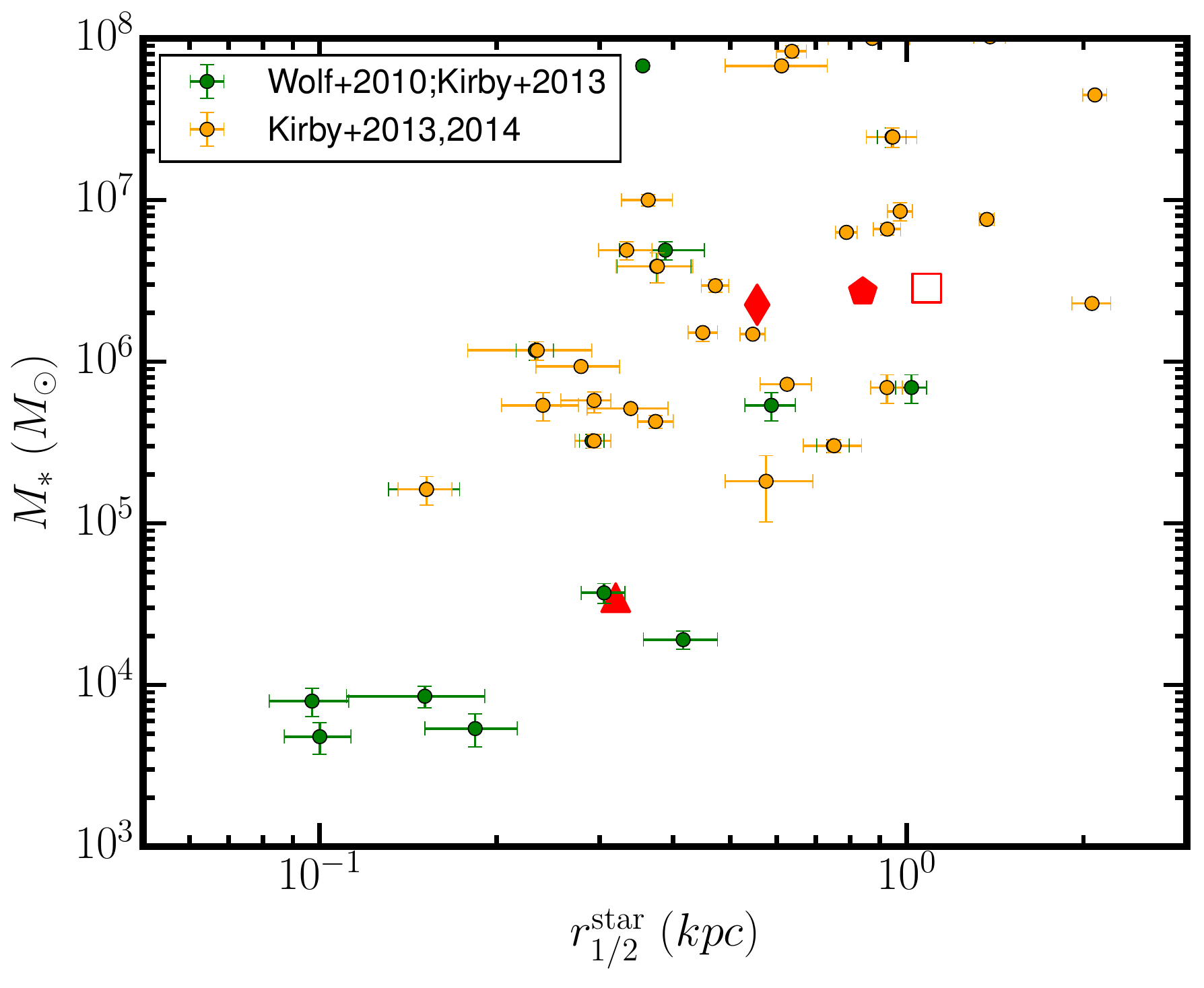}
\includegraphics[width=0.45\textwidth]{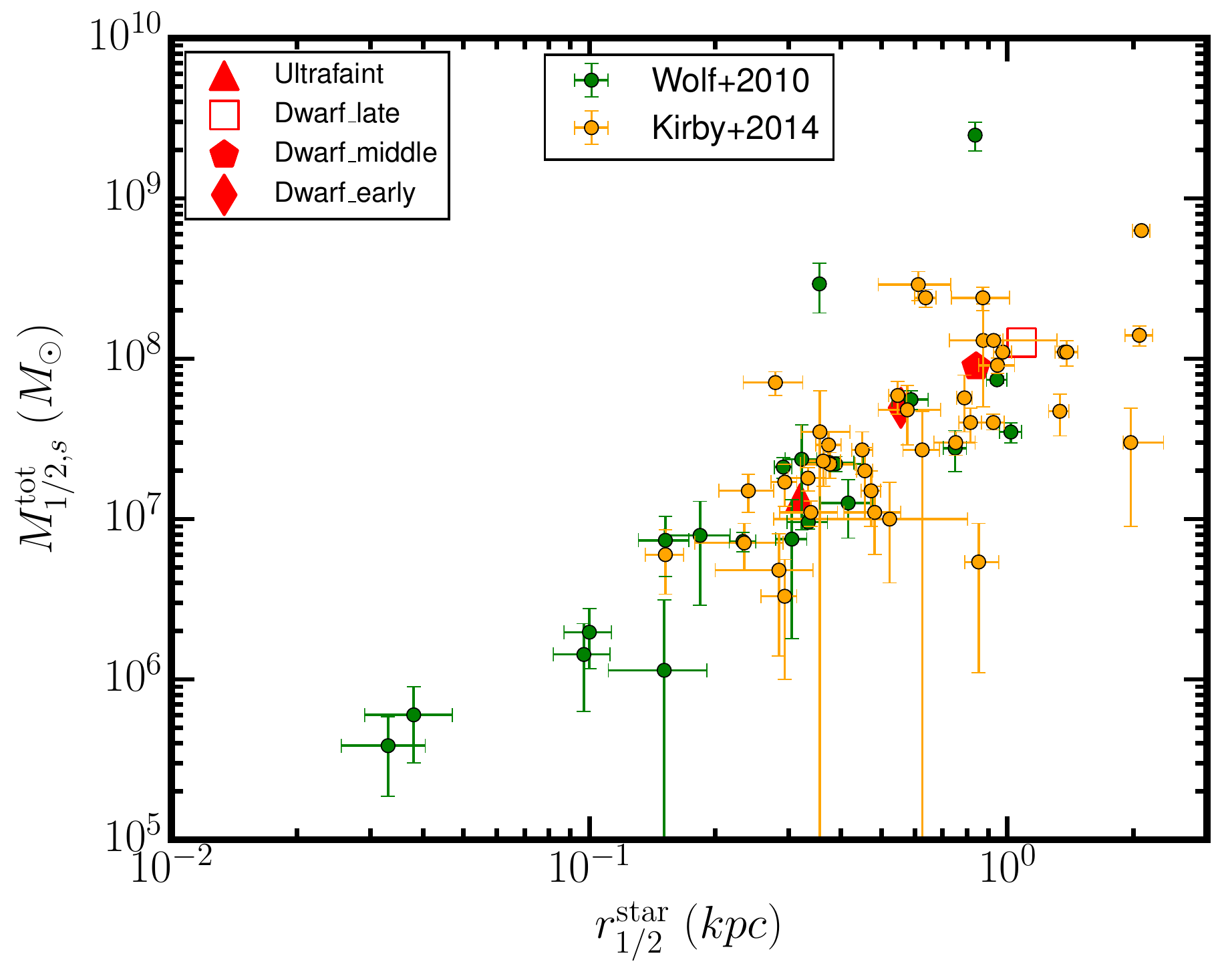}
\includegraphics[width=0.45\textwidth]{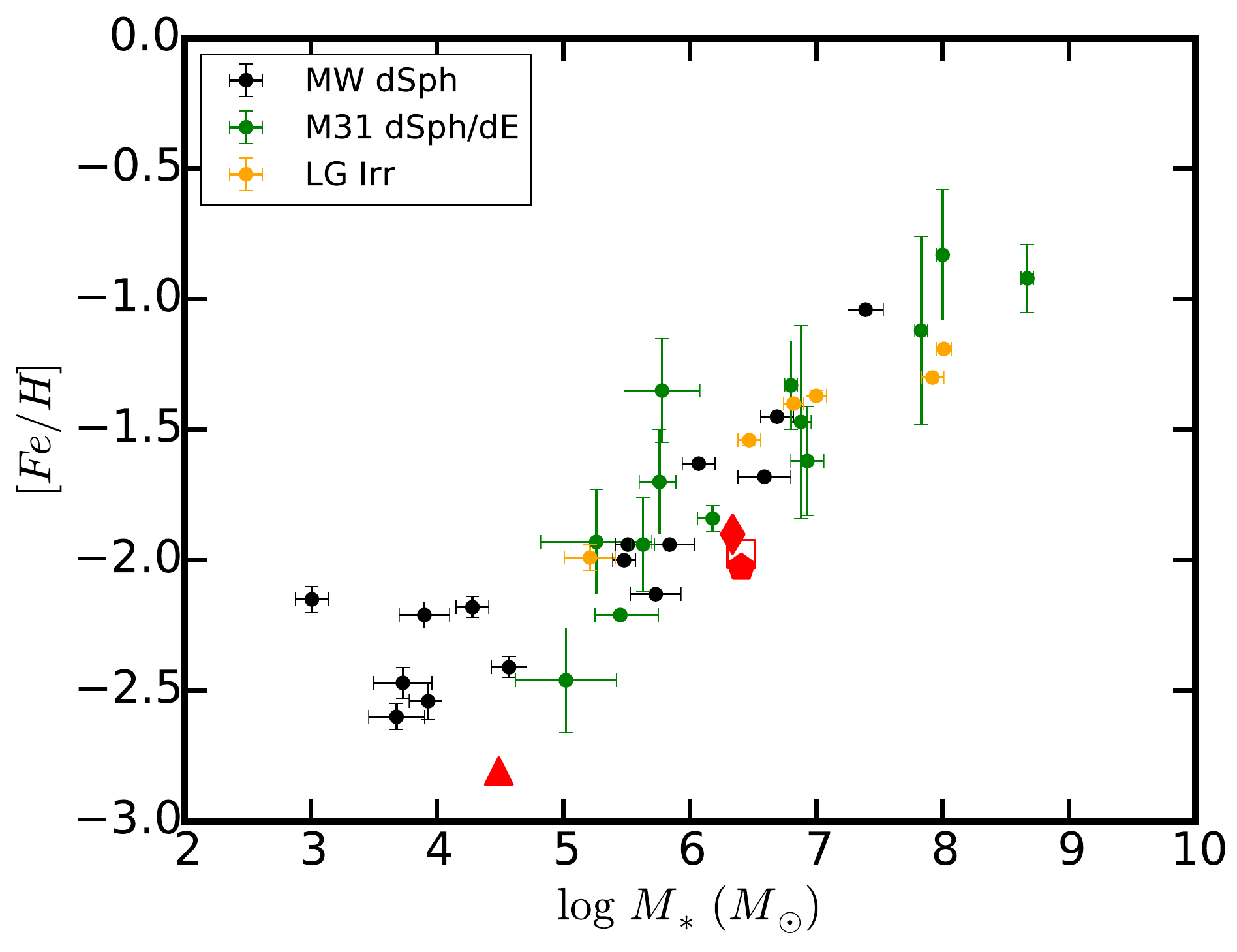}
\end{center}
\caption{All panels show the hydrodynamical runs presented in this work compared with different observations at $z=0$. Empty red square, filled red pentagon and filled red diamond stand for the Dwarf\_early, Dwarf\_middle and Dwarf\_late runs respectively. Upper left panel: Stellar mass versus effective stellar mass radius. Circles stand for observational data from \citep{Wolf:2010,Kirby:2013,Kirby:2014}. Upper right panel: The galaxy size (effective stellar mass radius) versus mass (total mass inside this radius) relation for the simulations presented in this work. In order to compare it with observations, we have plotted the data derived by \citet{Wolf:2010} (green circles) and \citet{Kirby:2014} (orange circles). Bottom panel: Stellar mass - stellar metallicity relation for the simulations presented in this work. Circles stand for the data compiled by \citet{Kirby:2013} for nearby dwarfs. See text for more details.}
\label{fig:propz0}
\end{figure*}

The upper left panel of Figure~\ref{fig:propz0} shows galaxy size, measured as the
half-stellar mass radius, versus the total stellar mass of the galaxy for our
simulated galaxies (red points).  The observed stellar size-mass relation seen
for Milky Way satellites (green) and Local Group field dwarfs (yellow) are shown
as data points \citep[taken from][who also compile data from the
literature]{Wolf:2010,Kirby:2013,Kirby:2014}. The upper right panel of
Figure~\ref{fig:propz0} shows the total mass within the half-stellar mass radius
vs. the half-stellar mass radius, again for our simulated galaxies compared to local
galaxies\footnote{When we do this comparison with observations we are assuming that
the half-stellar mass radius is equivalent to the half-light radius.}.
Finally, the bottom panel shows the stellar-mass metallicity
relation. Overall, the simulated galaxies are in good agreement with sizes, metallicities,
and total masses seen for galaxies of their stellar mass in the local universe.
For example, the Dwarf\_late and middle runs show a good agreement with
Fornax. Dwarf\_early's size and mass are close to Ursa Minor.

Table~\ref{tab:simsinfo} shows that Dwarf\_early is much more 
dark-matter-dominated within 500 pc than Dwarf\_late. 
This also holds when we look at the half-stellar mass radius of each galaxy,
instead of at a fixed physical value. Within this radius, Dwarf\_early 
has $M_{\rm tot}/\mstar \approx 44$ and 
$M_{\rm totl}/M_{\rm baryons} \approx 11$. By mass, stars are subdominant to gas
within the half-light radius by a factor of 2.8 for this dwarf. Within the half-mass stellar
radius, Dwarf\_late has $M_{\rm tot}/\mstar \approx 87$
and $M_{\rm tot}/M_{\rm baryons} \approx 5.4$
owing to a large reservoir of gas within the stellar half-mass
radius ($M_{\rm gas} = 16 \mstar$ within $\reobstar$).

\begin{table*}
\centering
\begin{footnotesize}
\begin{tabular}{l|cc|cccc}  
  \hline
  Parameter & Ultrafaint\_dm & Ultrafaint$^{1}$ & Dwarf\_dm  & Dwarf\_late & Dwarf\_middle & Dwarf\_early$^{2}$\\
   & (Collisionless) & (Hydro: Feed-V) & (Collisionless)  & (Hydro: Feed-M) & (Hydro: Feed-M-soft) & (Hydro: Feed-V) \\
  \hline
  1) $m_{\rm p}^{\rm dm}$ ($\msun$) & $1.5\times10^3$ & $1.26\times10^{3}$ & $1.5\times10^{3}$   & $1.26\times10^{3}$ & $1.26\times10^{3}$& $1.26\times10^{3}$\\
  2) $\epsilon_{\rm dm}$ (${\rm pc}$) & $28$ & $28$ & $35$  & $35$ &  $25$& $28$\\
  3) $m_{\rm p}^{\rm bar}$ ($\msun$) & -- & $2.54\times10^{2}$ & --  & $2.54\times10^{2}$ & $2.54\times10^{2}$& $2.54\times10^{2}$\\
  4) $\epsilon_{\rm gas}^{\rm min}$ (${\rm pc}$) & -- & $1.0$ & --  & $2.0$ & $25$& $2.8$\\
  \hline
  5) $\mvir$ $(\msun)$ &  $3.2\times 10^{9}$ &  $2.5\times 10^{9}$ & $9.2\times 10^{9}$ & $7.6\times 10^{9}$ & $7.7\times 10^{9}$ &$7.7\times 10^{9}$\\
  6) $\vmax$ $({\rm km/s})$ & $26$  & $22$ & $37$  &  $33$ & $33$ &$32.5$\\ 
  7) $\rvir$ $({\rm kpc})$ & $38$ & $35$ & $54$  & $51$ & $51$ &$51$\\
  8) $\lambda$ & $0.031$ &--& $0.0350$ & --& --& -- \\
  9) $V_{max}/V_{vir}$ & $1.35$ &--& $1.38$& --& --& -- \\
  10) $t_{50}$ $(Gyr)$& $1.43$  &--& $1.84$  & --& --& -- \\
  11) $\fbar \times (\Omega_{\rm m}  / \Omega_{\rm b}) $& -- & $0.024$ &  -- & $0.093$ & $0.074$ & $0.056$\\
  12) $\fgas$ & --  & $0.0049$ &  -- &  $0.018$ & $0.014$ &$0.011$\\
  13) $\fstar$ & -- & $0.00002$ & --  & $0.0004$ & $0.0003$ &$0.0003$\\
  14) $\mobstar$ $(\msun)$ & -- &  $2.1\times 10^{4}$ & --  & $2.8\times 10^{6}$ & $2.7\times 10^{6}$ & $2.2\times 10^{6}$\\
  15) $\reobstar$ $({\rm pc})$  & -- & $340$ & --  & $1100$ & $830$ & $550$\\
  16) $\FeH$  & -- & $-2.8$ & --   & $-2.0$ & $-2.0$& $-1.9$\\
  17) $\mtotTT$ $(\msun)$  & $2.2\times 10^{7}$  & $1.7\times 10^{7}$ & $3.4\times 10^{7}$  & $0.75\times 10^{7}$ & $1.3\times 10^{7}$& $1.9\times 10^{7}$\\ 
  18) $\mdmTT$ $(\msun)$   & $1.6\times 10^{7}$  & $1.7\times 10^{7}$ & $2.6\times 10^{7}$  & $0.43\times 10^{7}$ & $0.80\times 10^{7}$ & $1.6\times 10^{7}$ \\
  19) $\mbarTT$ $(\msun)$ & -- & $1.2\times 10^{4}$ & --  & $3.2\times 10^{6}$ &$4.7\times 10^{6}$ & $2.2\times 10^{6}$\\
  20) $\mgasTT$ $(\msun)$ & -- & $6.9\times 10^{2}$ & --  & $3.1\times 10^{6}$ &$4.5\times 10^{6}$& $1.6\times 10^{6}$\\
  21) $\mstarTT$ $(\msun)$& -- & $1.9\times 10^{4}$ & --  & $5.4\times 10^{4}$ &$2.4\times 10^{5}$ & $6.3\times 10^{5}$\\
   \hline 
\end{tabular}
\end{footnotesize}
\caption{Simulations data. First column stand for the different parameters studied for each simulation. In Columns 2-7 results at $z=0$ for the simulations presented in this work are shown. 
Row 1: dark matter particle mass in the high resolution region in solar masses. 
Row 2: fixed gravitational softening used for the dark matter particles in physical parsecs. 
Row 3: baryon particle mass in the high resolution region in solar masses.
Row 4: minimum baryonic force softening in parsecs (minimum SPH smoothing lengths are comparable or smaller). Recall that force softenings are adaptive (mass resolution is fixed).
Row 5: virial mass in solar masses defined at the overdensity at which the spherical top hat model predicts virialization \citep{Bryan:1998}.
Row 6: maximum circular velocity in $km/s$.
Row 7: virial radius in kiloparsecs.
Row 8: halo spin \citep{Bullock:2001} definition. See Appendix~\ref{app:dmevol} for more details.
Row 9: halo concentration.  See Appendix~\ref{app:dmevol} for more details.
Row 10: halo formation time. See Appendix~\ref{app:dmevol} for more details.
Row 11: virial baryon fraction, i.e., baryon mass inside the virial radius over the virial mass, divided by the cosmic baryon fraction.
Row 12: virial gas fraction, i.e., gas mass inside the virial radius over the virial mass.
Row 13: virial stellar fraction, i.e., stellar mass inside the virial radius over the virial mass.
Row 14: stellar mass in solar masses. This is the stellar mass of the central galaxy.
Row 15: effective stellar mass radius, i.e., half stellar mass radius in kiloparsecs.
Row 16: stellar iron over hydrogen ratio. Mass weighted iron over hydrogen ratio for the dwarf stellar mass component. 
Row 17: total mass inside $500$ parsec in solar masses.
Row 18: dark matter mass inside $500$ parsec in solar masses.
Row 19: baryon mass inside $500$ parsec in solar masses.
Row 20: gas mass inside $500$ parsec in solar masses.
Row 21: stellar mass inside $500$ parsec in solar masses.\newline
$^{1}$ simulation presented in \citet{Hopkins:2014} as m09.\newline
$^{2}$ simulation presented in \citet{Hopkins:2014} as m10.}
\label{tab:simsinfo}
\end{table*}

\subsection{Star Formation Histories}

\begin{figure*} 
\begin{center}
\includegraphics[width=0.45\textwidth]{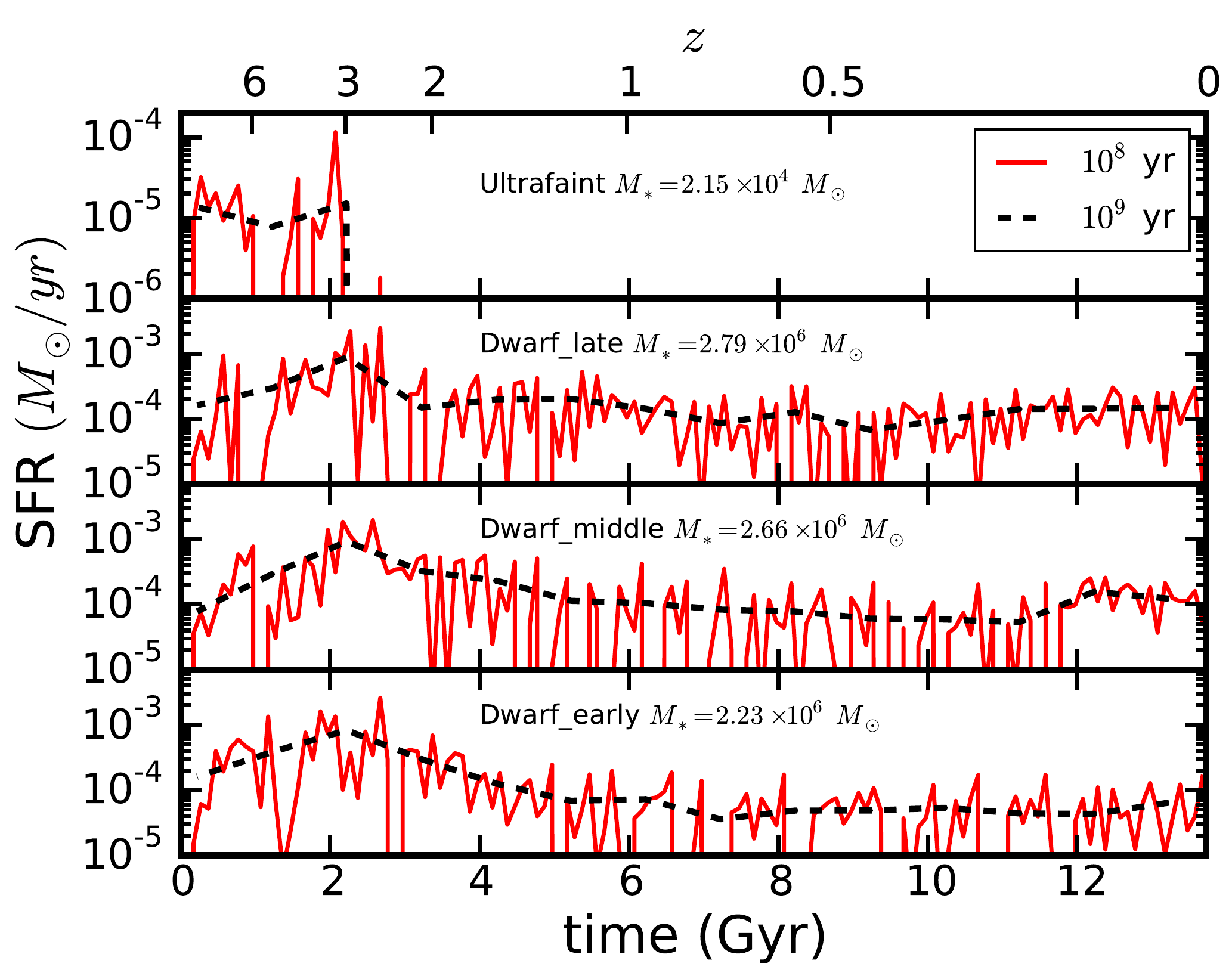}
\includegraphics[width=0.45\textwidth]{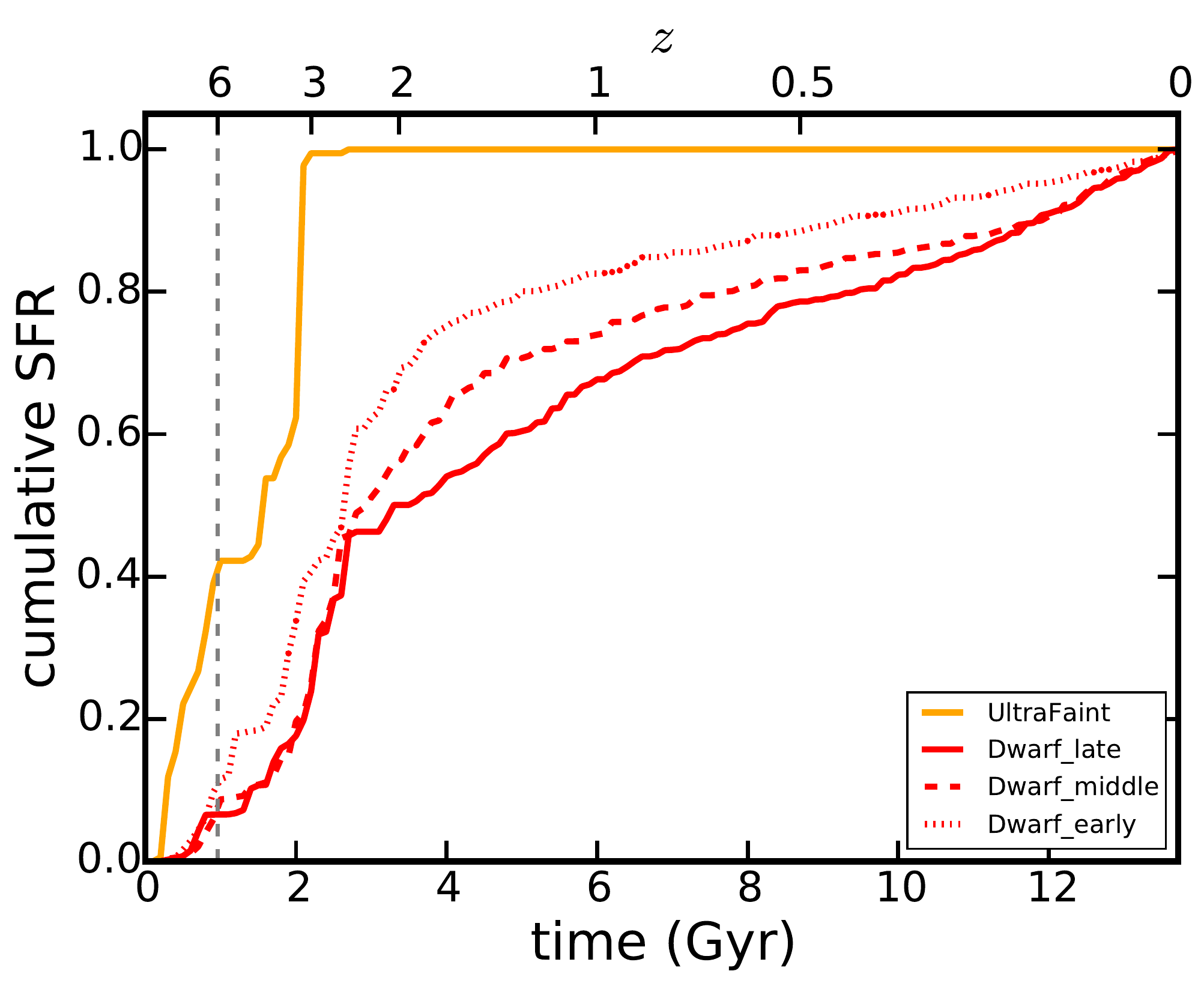}
\end{center}
\caption{Left panel: Star formation rates for the hydrodynamical runs
  presented in this work obtained using two different time bins: $10^{8}$ yr
  (full red lines) and $10^{9}$ yr (dashed black lines). Right panel: Normalized
  cumulative SFR history for all the hydrodynamical runs presented in this
  work. Notice the difference of the star formation histories between the three
  high resolution runs. The vertical grey dashed line marks the reionization
  redshift assumed in the simulation. Simulated star formation histories are
  very similar to some observed ones (e.g.,
  \citealt{Skillman:2014,Weisz:2014}). See text for more details.}
\label{fig:sfrnorm}
\end{figure*}

\begin{figure} 
\begin{center}
\includegraphics[width=0.45\textwidth]{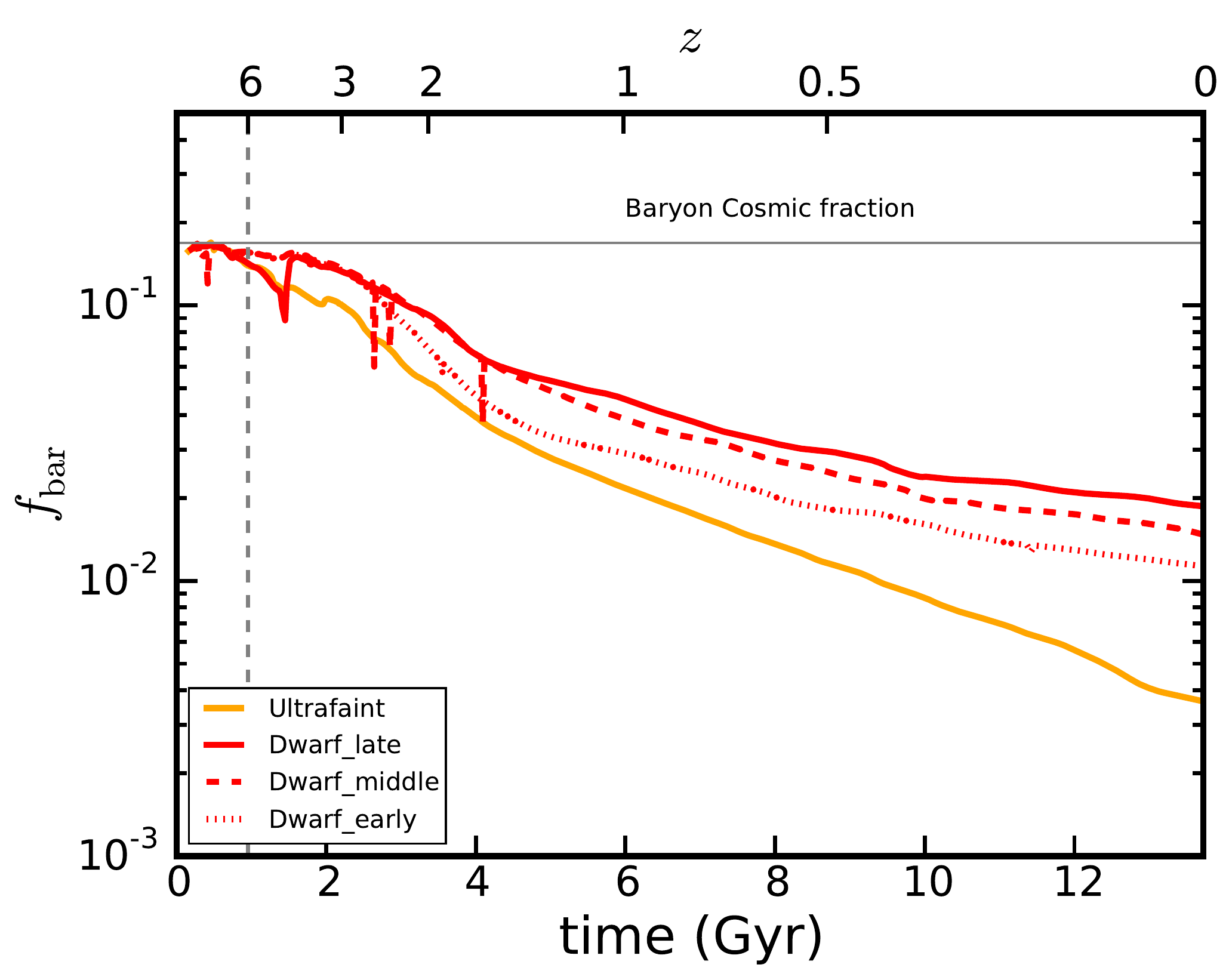}
\end{center}
\caption{Evolution of the virial baryon fraction for all the hydrodynamical runs presented in this work. In all cases, gas is slowly expelled out of the halo by the stellar feedback. Notice how the Dwarf runs with the highest SFR at early times (when the halo is still growing) end up having a lower baryon fraction. The Ultrafaint is further affected by the UV background, which prevents accretion of gas after $z\sim 6$.}
\label{fig:fbar}
\end{figure}

While the simulated Dwarfs (early, middle, and late) all show similar $z=0$ stellar masses, they arrived at those final states via different paths. The Ultrafaint run, on the other hand, ends up with a stellar mass some two orders of magnitude smaller than any of the Dwarf runs, though it resides within a halo that is only $\sim 3$ times less massive.  In this subsection, we explore these differences by examining the star formation histories in some detail.

In Figure~\ref{fig:sfrnorm}, we present the star formation rates (left panel) and the normalized cumulative star formation histories (right panel) of  all four high resolution hydro runs. 
The Ultrafaint simulation (orange line) forms all of its stars before $z\sim2.5$. The galaxy shuts down at this redshift because of two main effects: 1) the UV background prevents fresh gas accretion after $z \sim 6$ and 2) stellar feedback acts to self-quench the system after the ionizing background turns on. The remaining cold gas found at $z=0$ is not able to reach high enough densities to generate stars. Also notice that even though the UV background starts acting at high redshift on the gas particles, there is still active star formation for about one billion years after this redshift. 

This is not unexpected: previous lower-resolution simulations predicted that reionization-induced UV heating is not strong enough to remove all of the gas from dwarf-sized halos \citep{Hoeft:2006}. However, these simulations were not able to resolve the star formation histories of these galaxies, so it was not clear if the remaining gas would be able to form stars. As the UV background effectiveness depends on the density of the gas, cold and dense gas is not affected. Moreover, this more efficient star formation period seems crucial in order to match the stellar metallicity ratios observed for low-mass dwarf galaxies \citep{Kirby:2013}.

Figure~\ref{fig:sfrnorm} also shows the star formation rates of the three Dwarf runs.  Dwarf\_early forms more than half of its stars prior to $z=2.5$ while Dwarf\_late maintains a fairly substantial star formation rate down to $z=0$.  The Dwarf\_middle star formation rate history stands just between the two.  All of the runs show bursty star formation histories on $\sim 100$ Myr timescales.

In all three of the Dwarf runs, the star formation histories show two different phases.  At the highest redshifts ($z>3$), total dark halo and stellar masses both grow efficiently (albeit with some offset).
This is the ``rapid assembly" phase \citep{Wechsler:2002}, before/during reionization, in which feedback, while able to eject some gas from the galaxy and provide some overall suppression and variability of the star formation, does not appear to dominate the gas dynamics (the central potential and mass of the halo grow on timescales comparable to the galaxy dynamical time).
But from $z\sim3$ onward, halo accretion rates slow down and feedback acts strongly. From this point on, there appears to be a steady-state SFR that can be considered constant with time when averaged over a Gyr scale (a bursty behavior emerges when smaller time bins are used). In this phase, the galaxy is able to cycle new material into a fountain and so maintain equilibrium. This ``quasi-equilibrium" SFR scales with the central potential of the galaxy \citep[see][]{Hopkins:2012}, as traced by quantities such as the central halo density or $\vmax$ (the maximum circular velocity), not the halo mass or virial velocity. 
The central potential depth increases only weakly over this time as the halo accretes material mostly on its outskirts. This low but constant SFR at low redshift is a key factor in shaping the final matter structure of the dwarf galaxy and will be discussed in detail in the next Section.

Observations of the star formation rate of dwarf galaxies
\citep{Tolstoy:2009,Skillman:2014,Weisz:2014,Brown:2014, Cole:2014} show a
relatively high dispersion for a fixed stellar mass, but all the histories of
our simulated galaxies seem realistic 
when compared with these
data.  In particular, our Ultrafaint galaxy is composed of uniformly old stars,
as observed in real ultra-faint dwarf satellites of the Milky Way \citep{Brown:2014},
perhaps making them fossils of reionization \citep{Ricotti:2005, Bovill:2011}.

Some insight into the extremely low efficiency of star formation in all of these
systems can be gained from examining the total baryon fraction vs. time within
their associated virial radii.  This is shown in Figure~\ref{fig:fbar}.
Specifically, the virial baryon fraction (baryon mass divided by total mass
inside the virial radius) begins declining in the Ultrafaint run from the moment
the UV background starts acting to reduce the amount of gas falling into the
halo.  This allows feedback to be more efficient in expelling the gas out of the
halo potential.  The Dwarf runs also begin to demonstrate a steady decline in
their baryon fractions, but only after $z \sim 2.5$.  This is the redshift when
the halo itself stops growing (the end of the ``rapid accretion phase").  Below
that redshift, star formation feedback steadily acts to expel gas.  The runs
with the higher star formation rates have a lower baryon fraction, though all
three of the Dwarf runs end up at a factor of $\sim 10-20$ below the cosmic mean
(horizontal gray line).  However, the stellar feedback in the Dwarf\_early simulation
has managed to expel a larger fraction of material from halo, slowing down late
time star formation.

One intriguing result from Figure 5 is that the overall baryon fraction decreases steadily, without global jumps that are tightly linked to the star formation rate (which varies substantially over $\sim 100$ Myr timescales).  Instead, the baryons slowly ``evaporate" out. This is in contrast with other studies \citep[][]{Sawala:2011,Simpson:2013} that show sharper jumps.  However,  the lower resolution used by \citet{Sawala:2011} could explain why their evolution is less smooth. \citet{Simpson:2013} used comparable resolution to this work but studied a lower mass system, $10^{9} \msun$, which could explain the much more drastic effect due to the UV background in the gas virial fraction that they found. 

In the next Section, we study how these differences in the star formation histories affect the matter distribution of the halo.

\begin{figure*} 
\begin{center}
\includegraphics[width=0.45\textwidth]{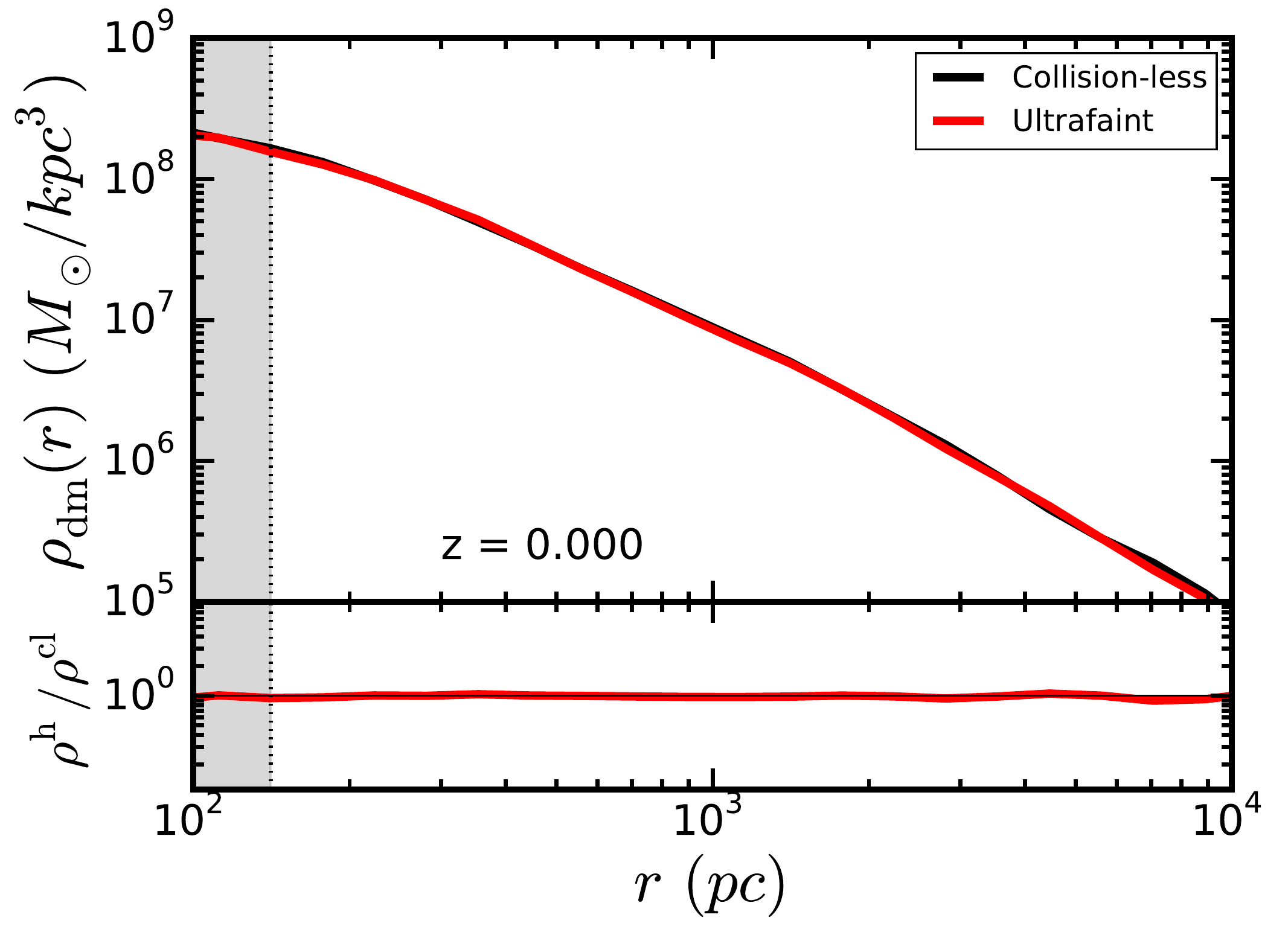}
\includegraphics[width=0.45\textwidth]{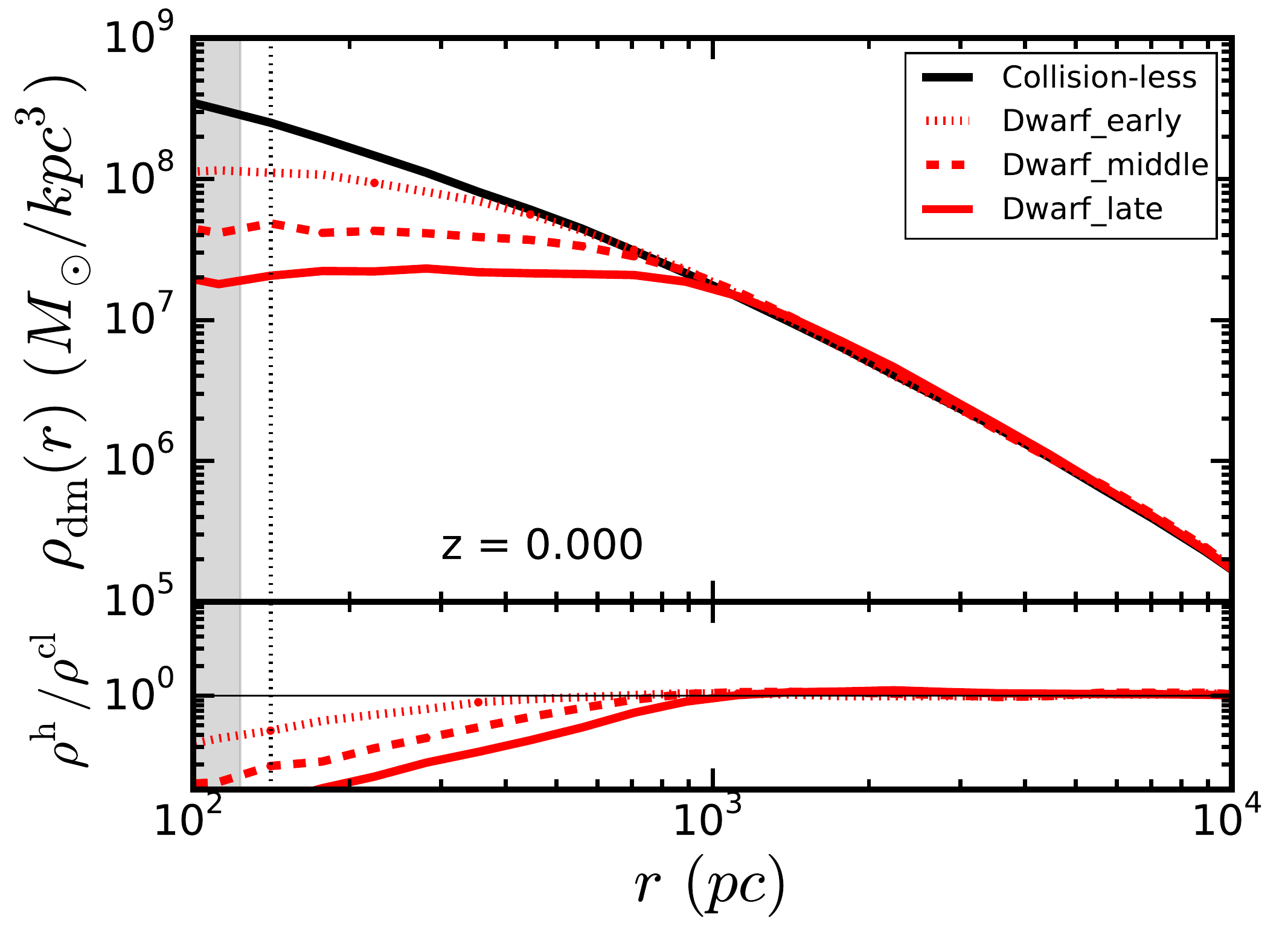}
\end{center}
\caption{Left: The dark matter density profile at $z=0$ for the collisionless (black line) and hydrodynamical (red line) runs of the $3\times 10^{9}$ $\msun$ halo. The ``collisionless" line has been converted to the effective dark matter density by accounting for the fact that a fraction $\Omega_{\rm b}/\Omega_{\rm m}$ of each particle is assumed to be baryonic in these runs.  The bottom panel shows the ratio between the two profiles. Right: The same for the Dwarf halo runs, where each hydrodynamical run is marked by a different style of red line.   Grey shaded area marks the region below the convergence radius defined using \citet{Power:2003} criteria for the colissionless run. The vertical black dotted line marks four times the dark matter gravitational softening used in the collisionless runs.  Note that the Dwarf\_late run has produced a large ($\sim 1$ kpc) constant-density core, while the Dwarf\_early  has a dark matter profile that is very similar to the dissipationless simulation for radii that are well converged.  The dark matter in the hydrodynamic Ultrafaint run is identical to that of the dissipationless case.}
\label{fig:dmprofz0}
\end{figure*}

\begin{figure*} 
\begin{center}
\includegraphics[width=0.45\textwidth]{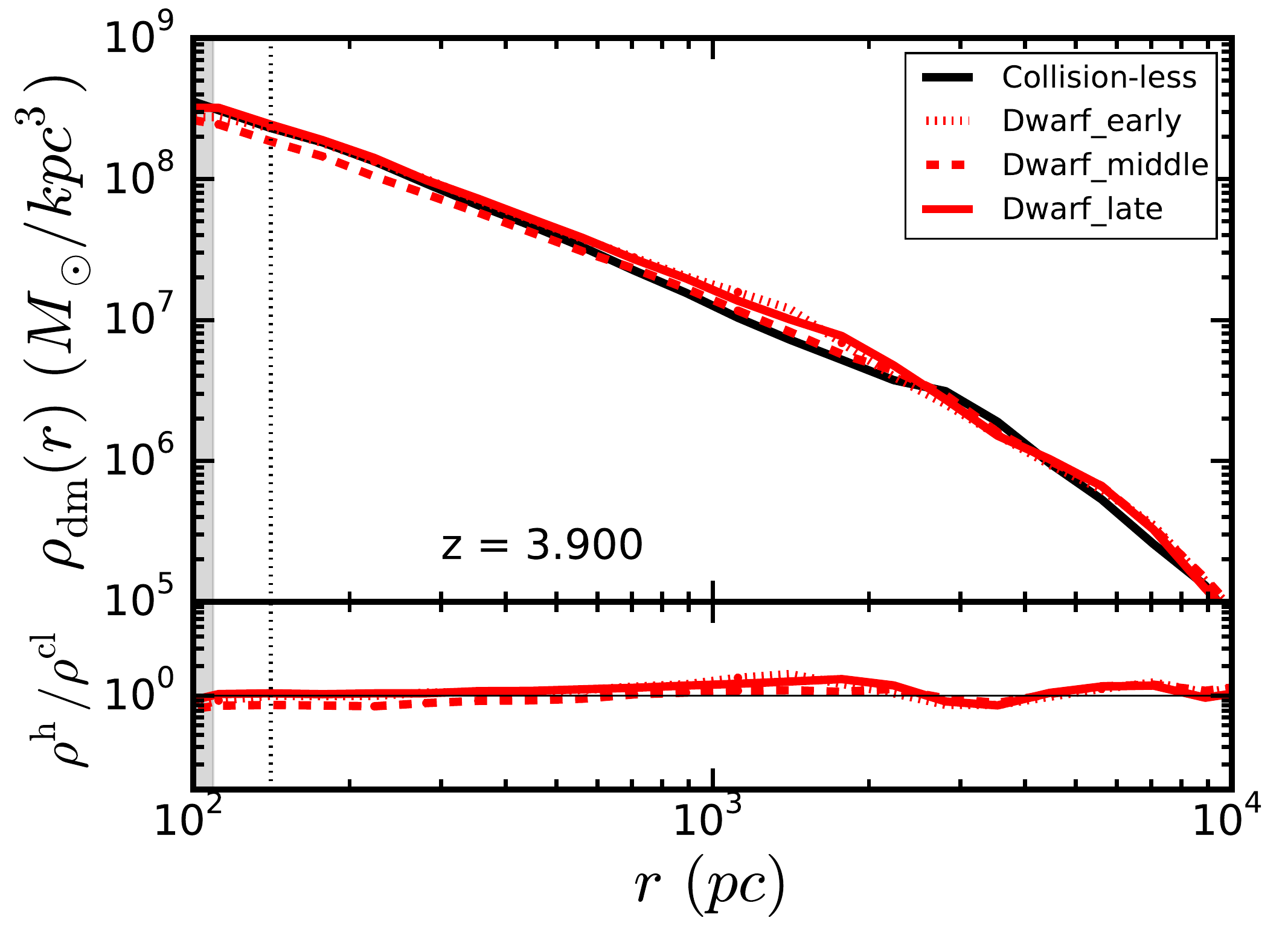}
\includegraphics[width=0.45\textwidth]{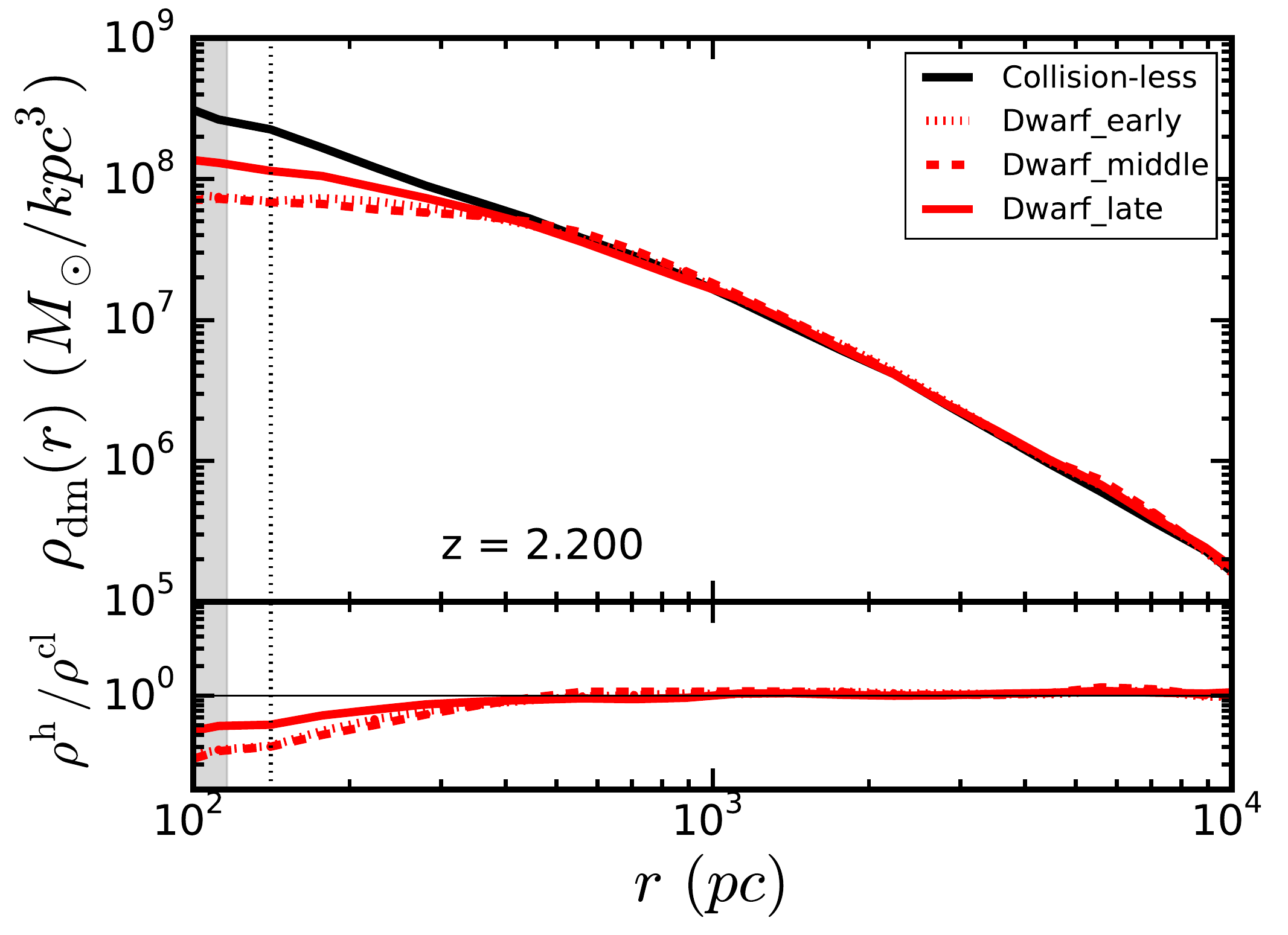}
\end{center}
\caption{Time variation in density profiles.  The dark matter density profile at $z=3.9$ (left figure) and $z=2.2$ (right figure) for the collisionless (black) and hydrodynamical (red lines) runs of the $1\times 10^{10}$ $\msun$ halo. Bottom panel shows the ratio between the two profiles. The vertical black dotted line marks four times the dark matter gravitational softening used in the collisionless runs.  Grey shaded area marks the region below the convergence radius defined using \citet{Power:2003} criteria for the colissionless run. Note that, at $z=2.2$, Dwarf\_late has a higher central density than Dwarf\_early.  Late-time star formation in Dwarf\_late serves to reduce the dark matter halo's density in the center by a factor of  $\sim 5$ by $z=0$ (see right panel of Fig. 6), while Dwarf\_early has little star formation subsequent to $z=2.2$. Its density profile remains essentially unchanged from $z=2.2$ to $z=0$.}
\label{fig:dmprofzevol}
\end{figure*}

\section{Dark matter content and structure}
\label{sec:dmstruc}

In Figure~\ref{fig:dmprofz0}, we present the dark matter density profiles of the
hydrodynamical Ultrafaint (left) and Dwarf (right) runs compared with their equivalent
collisionless run at $z=0$.  The grey bands mark the regions where the
simulations are not fully converged according to the criterion of \citet{Power:2003}
computed for the dark matter only simulation. 
The Power radius is defined to be the radius where the
two-body relaxation time, $t_{\rm relax}$, becomes shorter than the age of the
universe $t_{\rm 0}$, where $t_{\rm relax}$ is determined by the number of particles 
and the average density of the enclosed region $\bar{\rho}$. Specifically, Power et al.
found that $t_{\rm relax} < 0.6t_{\rm 0}$ is the best criterion. \citet{Elbert:2015} have
recently confirmed that this criterion is accurate using zoom simulations of collisionless dwarf
halos at similar resolution to those we examine here.  The vertical black dotted
lines in Figure~\ref{fig:dmprofz0} mark four times the dark matter gravitational softening used in the
collisionless runs. We note that while radii larger than the Power radius should not suffer 
from two-body relaxation, the smallest radius where results in hydrodynamical simulations
are converged may be (significantly) larger.

The left panel of Figure~\ref{fig:dmprofz0} shows results for the Ultrafaint
simulations. In this case, there is no sign of a decrease in the dark matter density in
the hydrodynamical run; in fact the dark matter
profile matches perfectly with the collisionless run. 
In the Dwarf runs (right
panel), we observe that all hydrodynamical runs show varying levels of decrease
of the inner dark matter density when compared with their equivalent
collisionless run.  In particular, the Dwarf\_late run has produced a fairly large ($\sim 1$ kpc) constant density
core -- this is exactly the behavior needed to help alleviate the Too Big to Fail Problem \citep[see, e.g.][]{Elbert:2015,Governato:2015}
and that would be required to explain indications of cored profiles in low-mass galaxies in the Local Group
 \citep[][]{Donato:2009,Salucci:2012,Walker:2011,Amorisco:2014,Burkert:2015}.

 One common way to quantify core formation in halos is to measure the log-slope
 of the density profile $\alpha$ at 1-2\% of the virial radius.  The Dwarf halo in the
 dark-matter only run has $\alpha=-1.58$, while Dwarf\_early, Dwarf\_middle, and
 Dwarf\_late have $\alpha=-1.39$, $-0.88$, and $-0.27$, respectively.  The
 late-forming dwarf produces the shallowest profile and the largest core, while
 the early-forming dwarf produces the densest, cuspiest system.

 \begin{figure*} 
\begin{center}
\includegraphics[width=0.48\textwidth]{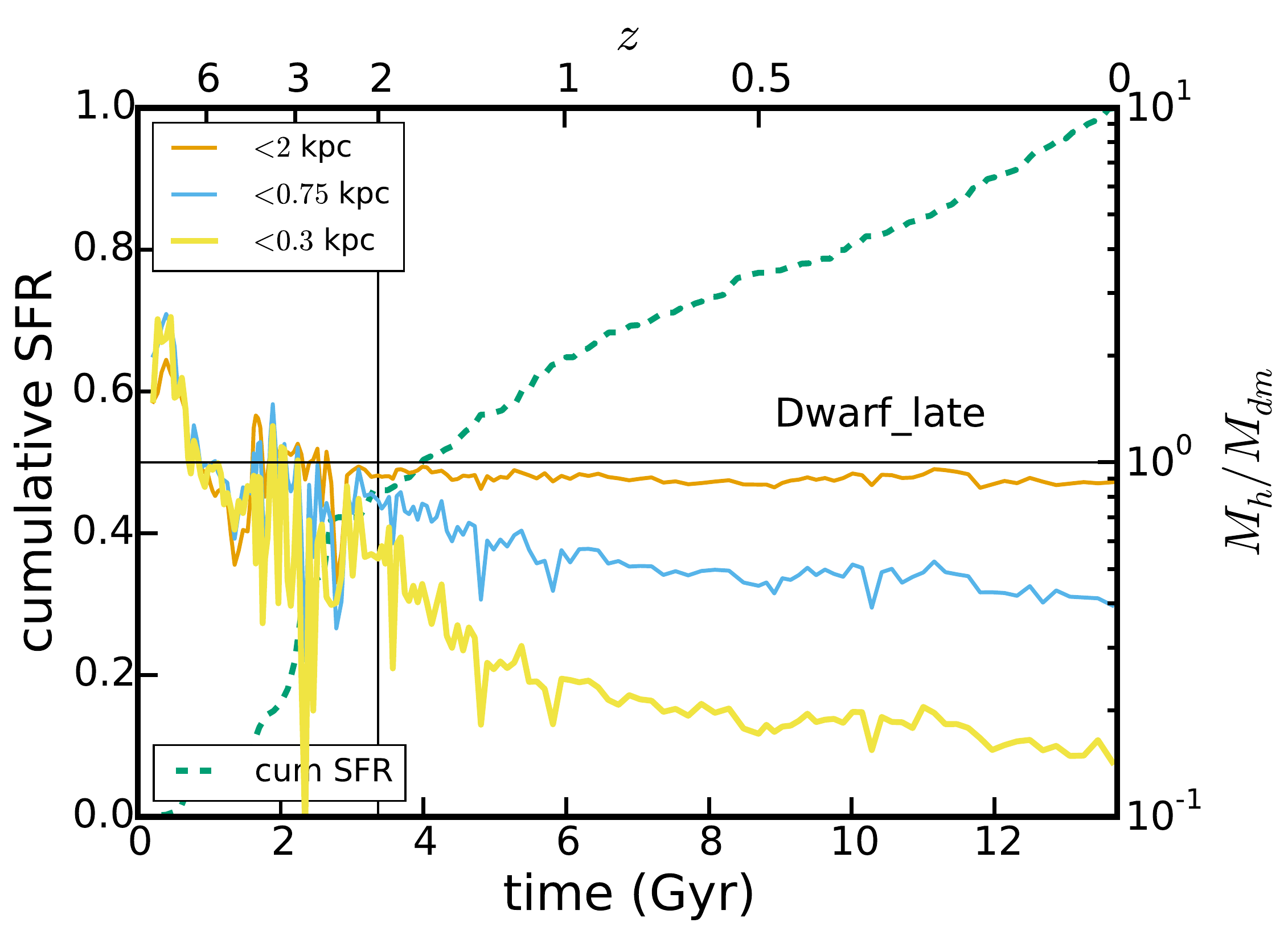} 
\includegraphics[width=0.48\textwidth]{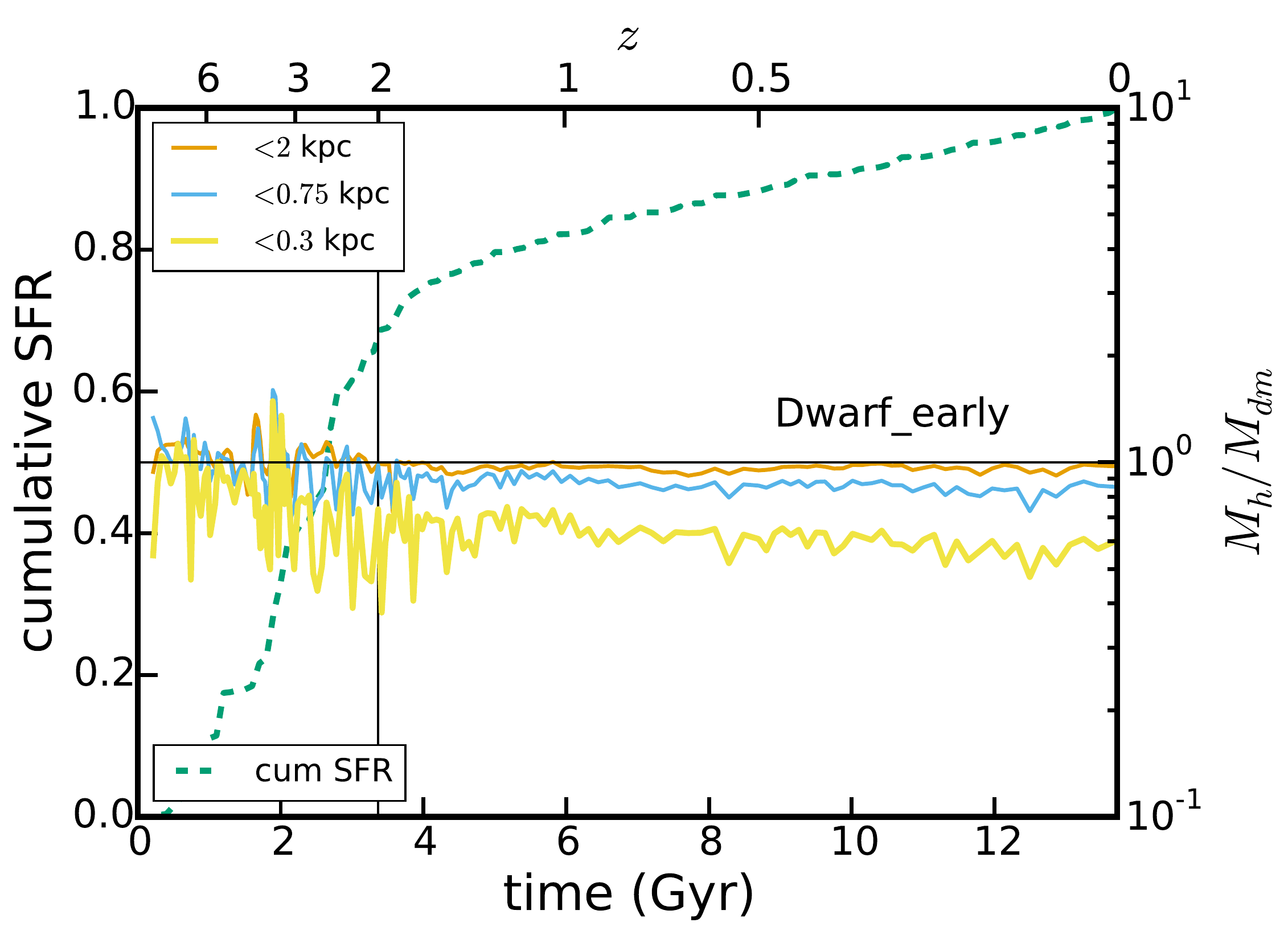}
\end{center}
\caption{Correlation between star formation history and core formation for Dwarf\_late (left) and Dwarf\_early (right).  The green dashed lines are the cumulative star formation histories for each run
normalized to its value at $z=0$.   The vertical line indicates the end-phase of the rapid-accretion epoch -- feedback after this time is most capable of forming stable cores that are not regrown by subsequent mergers.
 The solid lines show the integrated mass within different radii (as labeled) divided by the same
mass in the dissipationless runs -- values
less than unity indicate the formation of a lower density core than in the dark-matter only case.  We see that in the Dwarf\_late case (left), the core begins to form after the rapid accretion phase has ended ($z \sim 2$) and keeps growing slowly down to redshift $z=0$ as star formation continues.   Dwarf\_early, however, forms most of its stars prior to the quiescent accretion phase and a large core never takes hold.
}
\label{fig:massratio}
\end{figure*}
 
 Over time, in all three
 dwarf runs, we have observed clear correlations between core formation and
 star-formation events.  However, at early times, as the halos continue to
 accrete matter and experience central mergers, the cusps regrow regularly.
 During the early, rapid accretion phase, evolution in the density structure is
 fairly stochastic, with cores forming in response to blow-out events, and then
 becoming erased as cusps reform in response to mergers.
 Figure~\ref{fig:dmprofzevol} illustrates the formation of the dark matter core using two time
 steps.  Shown are the dark matter density profiles of the Dwarf runs at $z=3.9$
 (left) and $z=2.2$ (right). At $z=3.9$, very little star formation has occurred
 and the halo is experiencing very rapid growth and we see no decrease in core
 dark matter density compared to the collisionless run.  However, at $z=2.2$,
 there are some signs of a decrease in the central dark matter density in the
 hydro runs.  Interestingly, Dwarf\_late -- which has the largest core at $z=0$
 -- has the smallest core profile at $z=2.2$.  Dwarf\_early shows almost the
 opposite trend, owing to the fact that it has had more star formation by
 $z=2.2$ than the later forming dwarf.

 To further explore the evolution of the dark matter density with time,
 Figure~\ref{fig:massratio} compares the cumulative star formation history
 (dashed curves, normalized to unity at the present day) and the mass interior
 to radii of 0.3, 0.75, and 2 kpc relative to the collisionless run as a
 function of time for Dwarf\_late (left panel) and Dwarf\_early (right
 panel). In both cases, the early phases of galaxy formation ($z \ga 3$) result
 in fluctuations in the inner mass profiles of these galaxies \citep{Davis:2014}.
  After $z=3$, when
 the dark matter assembly of each halo is essentially complete, Dwarf\_early
 forms only a relatively small amount of stars. This results in at most a slight
 reduction in the inner dark matter mass (right panel). Dwarf\_late, however,
 forms more than 50\% of its stellar mass after $z=2$. Most of the density
 reduction also occurs after this phase, pointing to a link between the final
 densities of these objects and their late-time star formation histories. This
 is consistent with \citet{Laporte:2015}, who found that cusps can regrow after
 early core formation.

\begin{figure} 
\begin{center}
\includegraphics[width=0.45\textwidth]{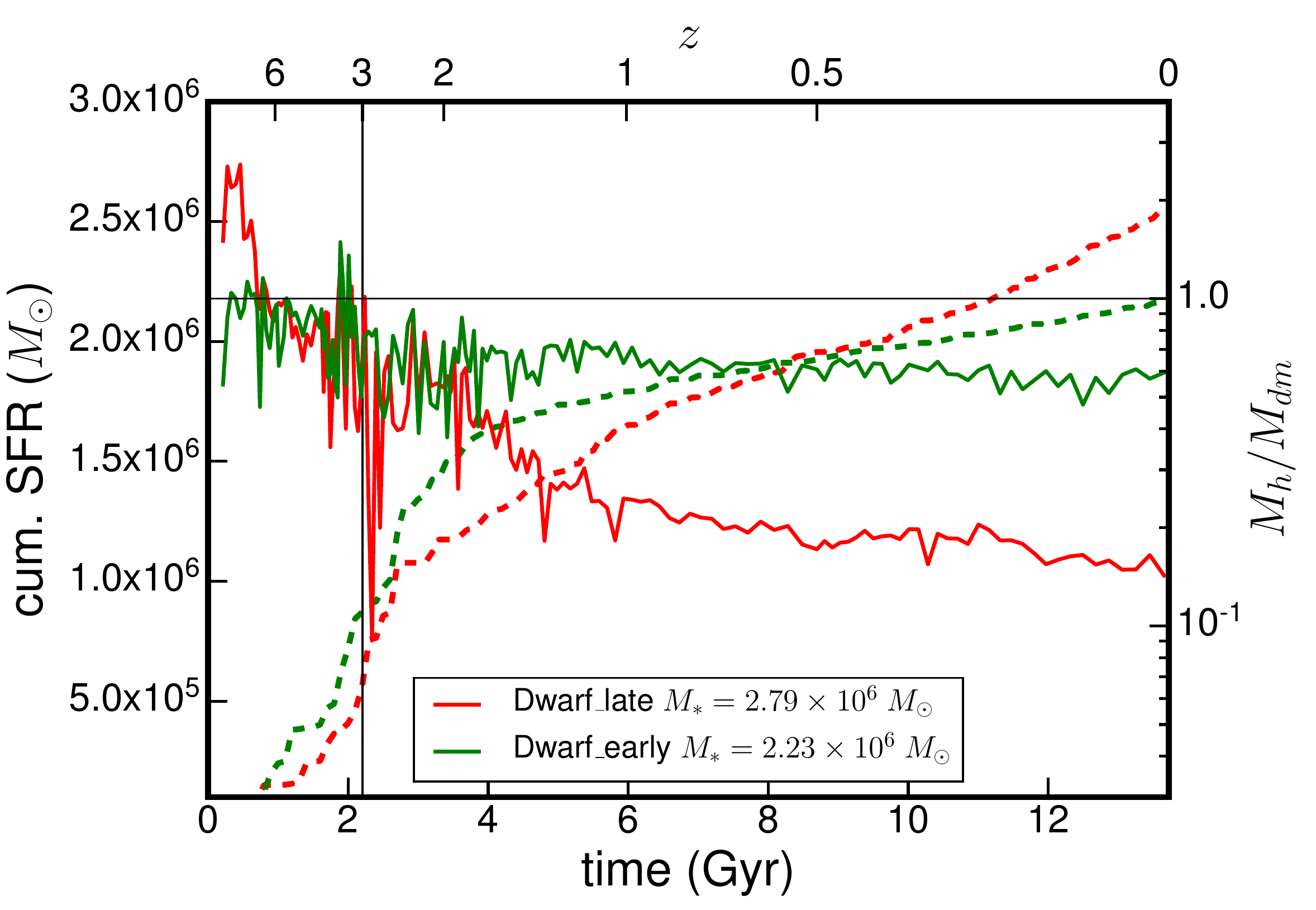}
\end{center}
\caption{The cumulative star formation history (dashed lines) and the dark matter mass ratio between the hydrodynamical run and the collisionless run at $0.3$ kpc (full lines) for the early and late forming Dwarf runs.}
\label{fig:massratioall}
\end{figure}

Figure~\ref{fig:massratioall} further illustrates the correlation between star
  formation history and core formation, now with the early and late runs on the
  same plot, and using dimensional star formation histories rather than
  normalized ones.  Specifically, the cumulative star formation histories of the
  Dwarf\_early (green dash) and Dwarf\_late (red dash) runs are shown along with
  the evolution of the ratio of dark matter enclosed within $0.3$ kpc for the
  hydrodynamic compared to the dark-matter only runs (solid lines).  It is clear
  that a higher star formation rate at late times (from $z\sim2$) produces a
  bigger decrease in the central dark matter density. Notice that although this
  difference in the star formation rate below $z=2$ produces very different
  cores, the difference in the total amount of stars at $z=0$ is minimal (see
  Figure~\ref{fig:sfrnorm} and Table~\ref{tab:simsinfo}). In concordance with
  this picture, lower resolution runs, which have slightly higher star formation
  rates at low redshift than their high resolution counterparts, show bigger
  cores in their dark matter distribution (see Appendix~\ref{app:convergence}).

It is likely that the relationship between {\em when} the stars form and core formation is
most important at this critical stellar mass / halo mass scale ($M_* \sim2-3\times 10^6 \msun$ within $\sim 10^{10} \msun$ halos).
Previous simulation efforts  \citep{Pontzen:2012,DiCintio:2014} have found that dark matter cores are usually not
created in galaxies with so few stars in halos below $\sim 10^{10}\msun$.  We suggest that at this critical mass scale, where
the energy from feedback sources is just at the edge of that required for core formation, small variations in star formation histories
can significantly alter the result.  Indeed, in our general analysis of the dark matter
properties in all FIRE runs, we find a similar transition around $10^{10}$
$\msun$ (Chan et al, in preparation).

\subsection{Energy considerations}
\label{ssec:energy}

Recently, there has been some discussion in the  literature about the energy requirements for the formation of a core in a dwarf galaxy halo \citep[see, for example][and references therein]{Penarrubia:2012,GarrisonKimmel:2013,Teyssier:2013} -- specifically, how many stars are required for there to be enough energy available to create a core?   At first comparison, the fact that the Dwarf\_late simulation was able to produce a sizable core 
with so few stars appears to be in contradiction to the results of Garrison-Kimmel et al. (2013), 
who suggested that cores this large are not energetically possible.  However, the host halo considered
in Garrison-Kimmel et al. (2013) is more concentrated than the one we consider here.  In order to explicitly check whether our results make sense energetically  we aim to compare the energy released in supernovae in our simulations to the difference in dark matter gravitational energy potential of the dwarf hydro runs and its collisionless version. 

The gravitational potential energy is defined as:
\begin{equation}
 U=-4\pi G \int^{r_{\rm max}}_{r_{\rm min}} r M(r) \rho(r) dr
\end{equation}
We computed this value numerically directly from the simulation data. We
considered $r_{\rm min}=0$ (using $r_{\rm min}=\epsilon_{\rm dm}$ or $r_{\rm
  min}=\epsilon_{\rm gas}^{\rm min}$ gives very similar results) and $r_{\rm
  max}=2\,{\rm kpc}$. We used this maximum radius because we are interested in the energy necessary to decrease the inner part of the density profile and from this point the dark matter profiles match almost exactly (see Figure~\ref{fig:dmprofz0}). More importantly, at larger radii differences between the gravitational energy potential can be significant just due to the exact position of the substructure between the collisionless and the hydrodynamic runs. Therefore this definition of potential energy for each run sets a lower energy limit on the amount of energy necessary to create a specific dark matter decrease in the inner part of a halo. We define  $\Delta U_{dm}$ as the difference between the potential energy of the hydrodynamical run and the potential energy of the collisionless run.

In order to obtain an estimate of the energy available from feedback we have
considered the energy available from SNe using the parameters from our
simulations: $E_{\rm tot}=(M_{\rm star}/m) \, f\, E_{\rm sn}$, where $E_{\rm
  sn}=1\times 10^{51}$ erg is the energy of one supernova, $f=0.0037$ is the
fraction of stars more massive than $8 \,\msun$ for a \citep{Kroupa:2002} IMF,
and $m=0.4\,\msun$ is the mean stellar mass. The stellar mass of the central
galaxy at $z=0$ is between $\sim 2.3-2.8 \times 10^{6} \msun$ however from
Figure~\ref{fig:massratio} we can see that the core starts to form below
$z\sim2$, therefore we have also considered the stellar mass produced since this
time until $z=0$.
We are considering just supernova energy but in principle, just taking
into account the energy, the contribution from photoionization and radiation pressure
could play a role in core creation. Although most of the energy in radiation just
escapes the system, there is in principle $\sim100$ times more energy in
radiation than in supernovas. Preliminary tests done in this regard by changing the
energy per supernova do not point towards a relevant role of these processes, 
at least as they are currently implemented in the code. However we leave
a more careful analysis to future work.

\begin{figure} 
\begin{center}
\includegraphics[width=0.45\textwidth]{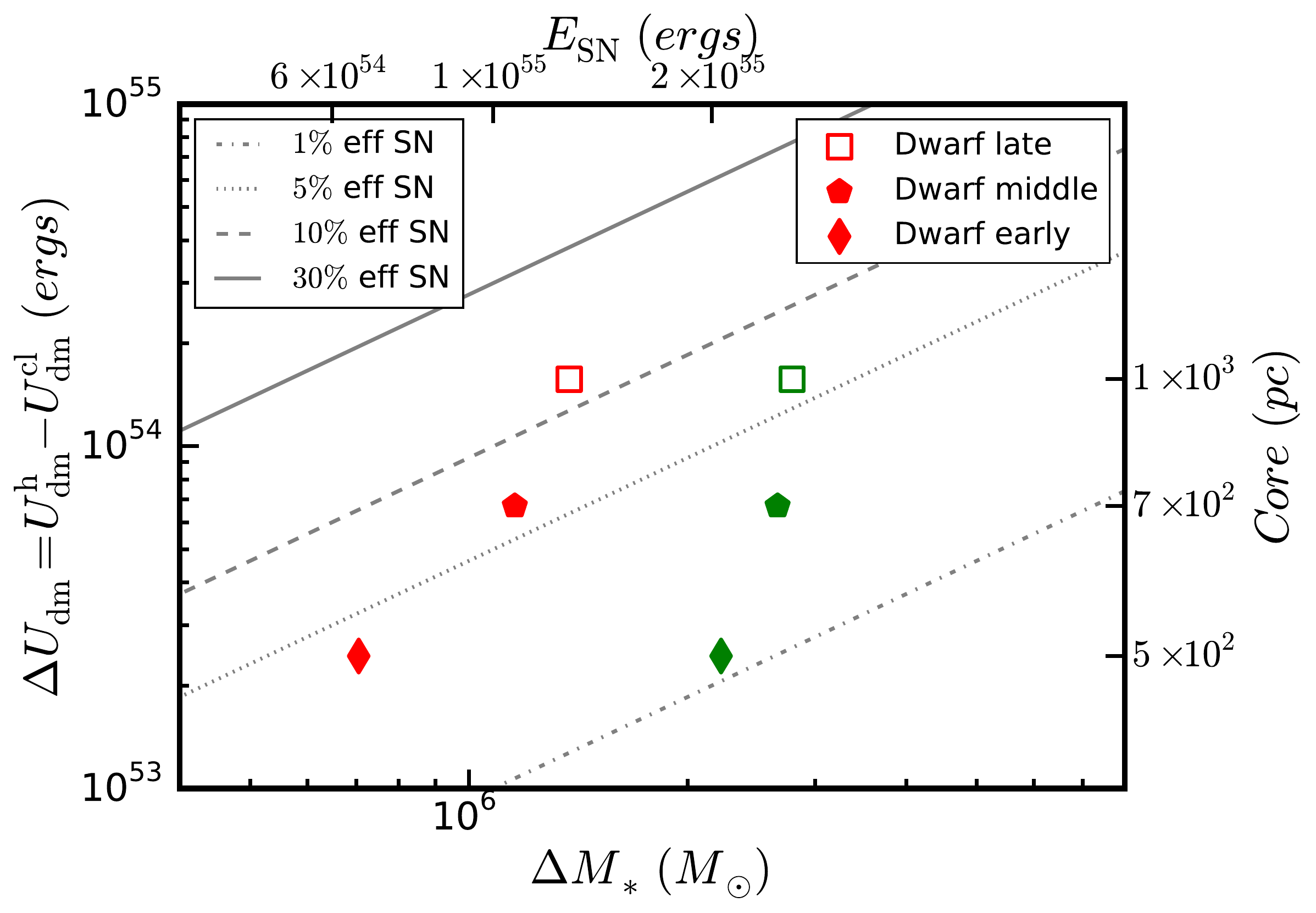}
\end{center}
\caption{Energy considerations in the formation of the dark matter cores in a $10^{10} \msun$ halo. The x-axis show the total stellar mass for each of the dwarf runs (green symbols) and the amount of stellar mass formed between redshift $z=2$ and $z=0$ (red symbols). The upper axis show the energy expelled by SN from this stellar population. The y-axis show the difference in potential energy between the hydrodynamical run and the collisionless run. The right y-axis show the size of the core created due to this difference of energy. See text for details.}
\label{fig:energycore}
\end{figure}

In Figure~\ref{fig:energycore}, we plot the potential energy difference between
the hydrodynamical runs and the collisionless run, $\Delta U_{\rm dm}$, versus
the total stellar mass (green symbols) and the stellar mass produced since $z=2$
down to $z=0$ (red points) for all the different Dwarf runs. The supernova
energy available from the stellar mass and the size of the core\footnote{We
  define the size of the core at the radius where the mass ratio between the
  hydrodynamical over the colissionless runs is $0.9$ (see
  Figure~\ref{fig:dmprofz0}).} linked with the difference in energy are also
marked in the Figure. Grey lines stand for different efficiencies of the
supernova (1\%, 5\%, 10\% and 30\%), i,e. the energy from supernova affecting
the dark matter. Our simulations indicate that the gravitational potential energy
of the halo has changed by an amount consistent with $\sim 5-10\%$ of the SNe energy
available for the Dwarf\_late run, while only $\sim 1 \%$ of the total SNe energy has been
effectively captured by the dark matter in the Dwarf\_early run, owing to the fact that many of these
supernovae exploded during the period of rapid accretion, when cusps were reforming. 
We find similar results when considering the lower resolution runs, including efficiencies,
as they have slightly bigger cores than their counterparts but a higher star
formation rates at lower redshift (see Appendix~\ref{app:convergence}).
Overall, these results suggest that large cores induced by star formation feedback 
will never appear in galaxies smaller than $\sim 10^{5} 
\msun$ (in $\sim 2-3 \times 10^{9} \msun$ or bigger halos), owing to the inefficiency of SNe energy coupling to the dark matter.

\section{Summary and Conclusions}
\label{sec:conc}

We have performed several high-resolution zoom-in hydrodynamical simulations of
an ultrafaint galaxy halo ($3\times 10^{9}$ $\msun$) and a dwarf galaxy halo
($1\times 10^{10}$ $\msun$). Our simulations include all major sources of
stellar feedback, implemented directly from stellar evolution
calculations. Without parameter tuning, the code reproduces a relation
between galaxy stellar mass and halo mass that is consistent with observations.
Moreover, we find that global properties of these simulated halos -- including
their characteristic sizes, metallicities and gas contents -- are well-matched
to observed galaxies of similar stellar mass.  These global properties
describing the simulated dwarfs are robust to changes in force and mass
resolution. Furthermore, the feedback models and the outflows they generate are
inherently multi-phase, matching observations. The predictive nature of our
galaxy formation model is particularly important, as the model does not contain
ad hoc numerical solutions adopted by other models, e.g. cooling shut-offs or
prescribed wind properties, that contain adjustable parameters. 
The mass scale of our simulated dwarfs -- $\mstar \sim 2 \times 10^{6}\,\msun$
-- is particularly relevant because previous models able to generate cores have
usually formed almost an order of magnitude more stars in such halos 
($\mvir =10^{10}\,\msun$). Such
galaxies are too massive in terms of the number of stars given their halo mass
and therefore cannot be typical, given observed galaxy counts around the Milky
Way and generic predictions from \lcdm\ simulations \citep{GarrisonKimmel:2014b,
Brook:2014}.

Our models show a slow but continuous decrease of the baryonic mass inside 
the virial radius after $z\sim6$. The UV background, in concert with star
formation feedback, plays a fundamental role in regulating star formation
in low-mass systems and appears to be the driving factor in suppressing 
gas accretion in our ultrafaint run. 
However, for the halo masses studied in this work, the UV background
does not shut down star formation immediately because it is not efficient in
heating the high density gas in the center of these halos. The simulated
ultra-faint ($\mvir =3 \times 10^9 \,\msun$) continues forming stars for $\sim
2$ Gyr following reionization (at which time it runs out of cold gas; such an
object would be a counterpart in the field to known ultra-faint satellites of
the Milky Way), while the more massive dwarfs continue to form stars to
$z=0$. This may indicate a transition from lower-mass objects that are incapable
of acquiring cold gas after reionization to dwarfs at this mass scale that can
continue to accrete fuel for subsequent star formation.

We have also studied, in detail, the dark matter distribution of these
halos. The simulated dwarfs ($\mstar \sim 2\times 10^{6}\,\msun$) have a variety
of density profiles, ranging from a small modification of the equivalent
dark-matter-only simulation to a substantial (kpc-scale) core. The simulated
ultra-faint galaxy ($\mstar \sim 2 \times 10^{4}\,\msun$) does not form enough
stars to modify its dark matter halo at all, providing further support to the
idea that there is a critical mass below which core formation caused by stellar
feedback is energetically impossible (see, e.g., 
\citealt{Governato:2012,GarrisonKimmel:2013, Madau:2014}).
Our results indicate that stellar mass is
not the only parameter in core creation, however. The creation of dark matter
cores is linked with \textit{late-time} star formation properties, as only the
system with significant late-time star formation forms a sizeable core. The
galaxy that forms most of its stars at early time is able to create a core
temporarily, but subsequent dark matter accretion and mergers 
and the lack of strong star formation erase this core,
leaving a cuspier profile. The difference in density at $300\,{\rm pc}$
between these two extreme cases is a factor of $\approx 4$. A related point is
that the formation of stable dark matter cores is a continuous process, not
instantaneous, and that the creation of significant cores in dwarf galaxies
does \textit{not} appear to be an inevitable outcome in models with bursty star formation
histories.

A question that remains unclear is whether these cored systems can avoid regenerating
a density cusp once they merge with smaller, cuspier, haloes
\citep{Laporte:2015}. The late-time merger history of dwarfs can vary
significantly \citep[e.g.,][]{Deason:2014}, meaning it is imperative to simulate
a statistical sample of halos at a given mass to fully understand trends in core
creation or cusp regrowth (Fitts et al., in preparation). It will also be
imperative to test this scenario at different halo masses (Chan et al., in
preparation), as many models predict a core formation efficiency that varies
with the halo mass (e.g., \citealt{DiCintio:2014}).
We  have  not  considered the effects stripping from
ram  pressure  and  tides,  that  may  be  important  for  some
Milky  Way  subhalos \citep{Read:2006,Zolotov:2012,Brooks:2014}.
However, the central prediction
coming from our simulations is observationally testable: \textit{the presence of
cores in galaxies with stellar masses of $\sim 10^{6}-10^{7}\,\msun$ requires
substantial \textbf{late time} star formation}.

\section*{Acknowledgments}
This work used computational resources granted by NASA Advanced Supercomputing
(NAS) Division, NASA Center for Climate Simulation, Teragrid and by the Extreme
Science and Engineering Discovery Environment (XSEDE), which is supported by
National Science Foundation grant number OCI-1053575.  JO and JSB were supported
by NSF grant AST-1009999 and NASA grant NNX09AG01G. JO thanks the
financial support of the Fulbright/MICINN Program.
JO also thanks the pynbody team for making this software publicly available.
DK was supported by a Hellman Fellowship and NSF grant AST-1412153.
CAFG was supported by NSF through grant AST-1412836, by NASA through grant NNX15AB22G, and by Northwestern University funds.


\bibliography{ms}
\bibliographystyle{mn2e}

\appendix
\section{Dark matter properties and evolution in the colissionless run}
\label{app:dmevol}

\begin{figure} 
 \includegraphics[width=0.49\textwidth]{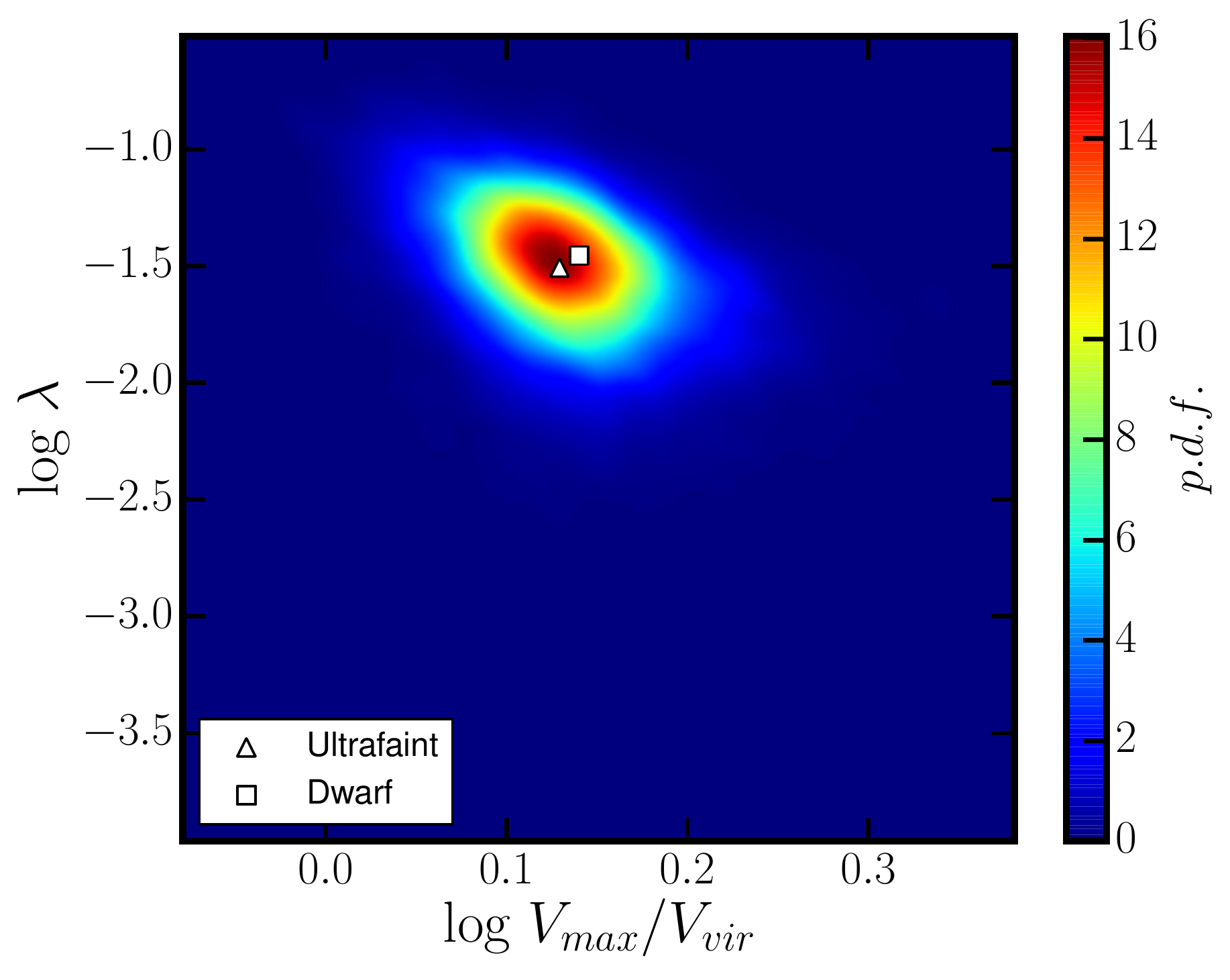}
 \includegraphics[width=0.49\textwidth]{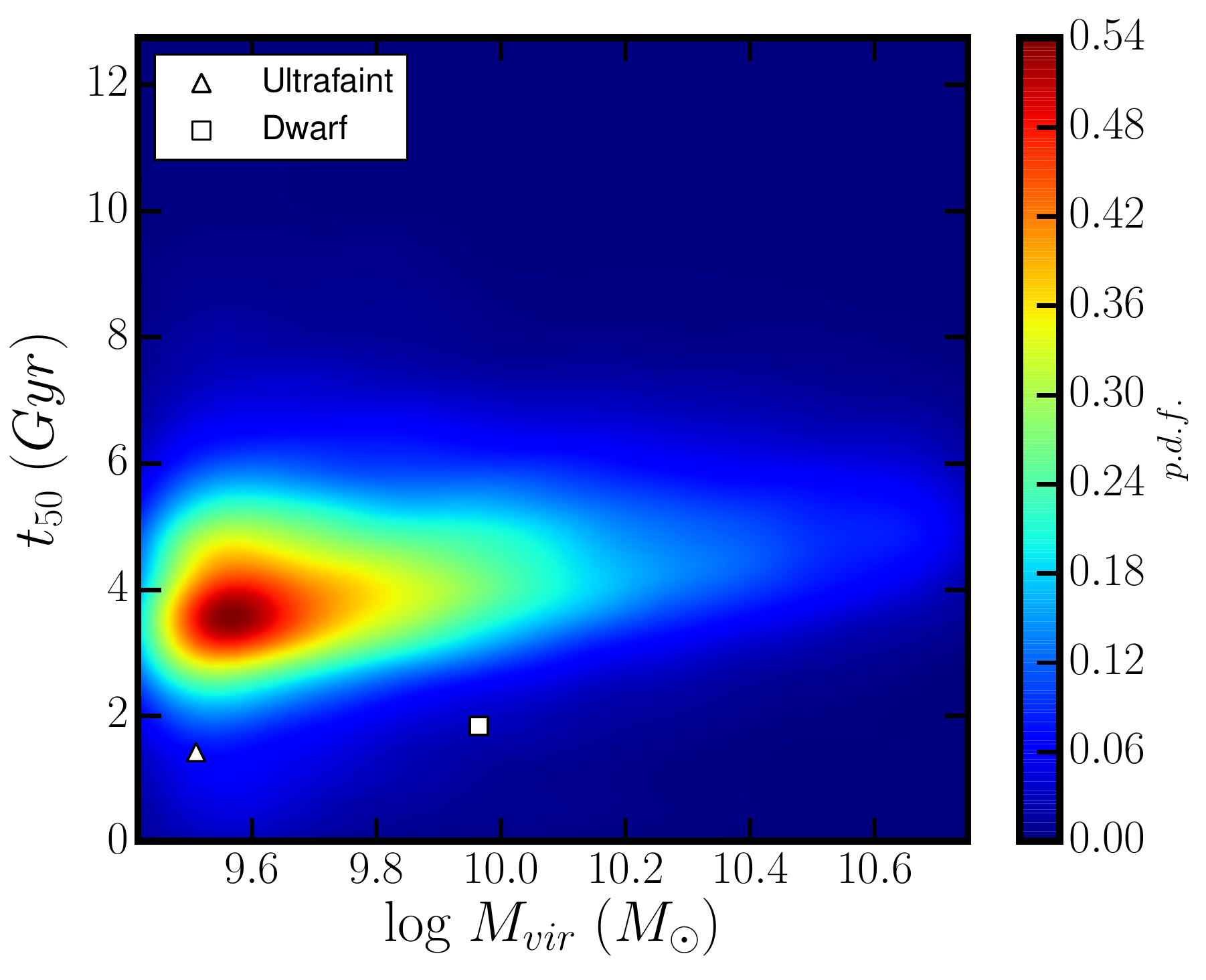}
 \caption{Selecting  the  sample. Color  map  shows  the  probability density function of the dwarf halo sample in the $L_{box}= 35$ $Mpc$ collisionless simulation ($512^3$ resolution). The white triangle and white square stand for the Ultrafaint\_dm and Dwarf\_dm runs respectively. Upper panel shows concentration versus halo spin. Lower panel shows the virial mass versus halo formation time. Specific values of these parameters can be found in Table~\ref{tab:simsinfo}.}
 \label{fig:dmsample}
\end{figure}

To choose the specific dwarf galaxy halos to re-simulate we rely on colissionless simulations. We have taken into account two things: first we wanted them to be cheap in terms of cpu cost and second we wanted them to be representative of the dwarf galaxy halo population.  We point to \citet{Onorbe:2014} for a full description of the method. Here we just want to show how the properties of our selected halos compare with a realistic sample of dwarf galaxy halos. To generate this sample we run a $L_{box}=35$ $Mpc$ collisionless simulation ($512^3$ simulations). Figure~\ref{fig:dmsample} show the spin ($\lambda$), concentration ($V_{max}/V_{vir}$), halo formation time ($t_{50}$) and virial mass distributions for all the main halos in this simulation (so excluding subhalos) with virial masses between $3\times 10^{9}$ $M_{\odot}$ and $3\times 10^{10}$ $M_{\odot}$. The mass bin sample includes around $\sim15000$ halos.
Halo spin parameters were calculated using \citet{Bullock:2001} definition. In order to estimate the time of formation for each halo, we followed the approach described in \citet{Wechsler:2002}. We fit the halo accretion histories obtained from the merger trees to a exponential form that depends on one parameter. The halo formation time $t_{50}$ is calculated at the time when the halo reached half of its total mass.  The chosen parameters for our ultrafaint and dwarf initial conditions are plotted as a white triangle and a white square respectively. The exact values can be found in Table~\ref{tab:simsinfo}. This Figure show that the re-simulated halos picked from our $L_{box}=7$ $Mpc$ box have very typical values of spin and concentration. The reason why our formation time is a bit lower than the standard value is a combination of three factors. First we preferred to avoid systems with late major mergers events which also helps to reduce the cpu cost of the simulation \citep{Onorbe:2014}. The dwarf halo sample from a smaller box is biased towards smaller formation times so there were a smaller range of possible halos to pick which fulfill all our desired criteria. Finally, the circular velocity profile of the Dwarf\_dm simulation, which can be found in \citet{Elbert:2015} (left panel), shows that the halo has too high density to match the circular velocity observations of Local Field dwarf galaxies. This makes it a suitable candidate to study the Too Big To Fail problem.

\begin{figure} 
 \includegraphics[width=0.49\textwidth]{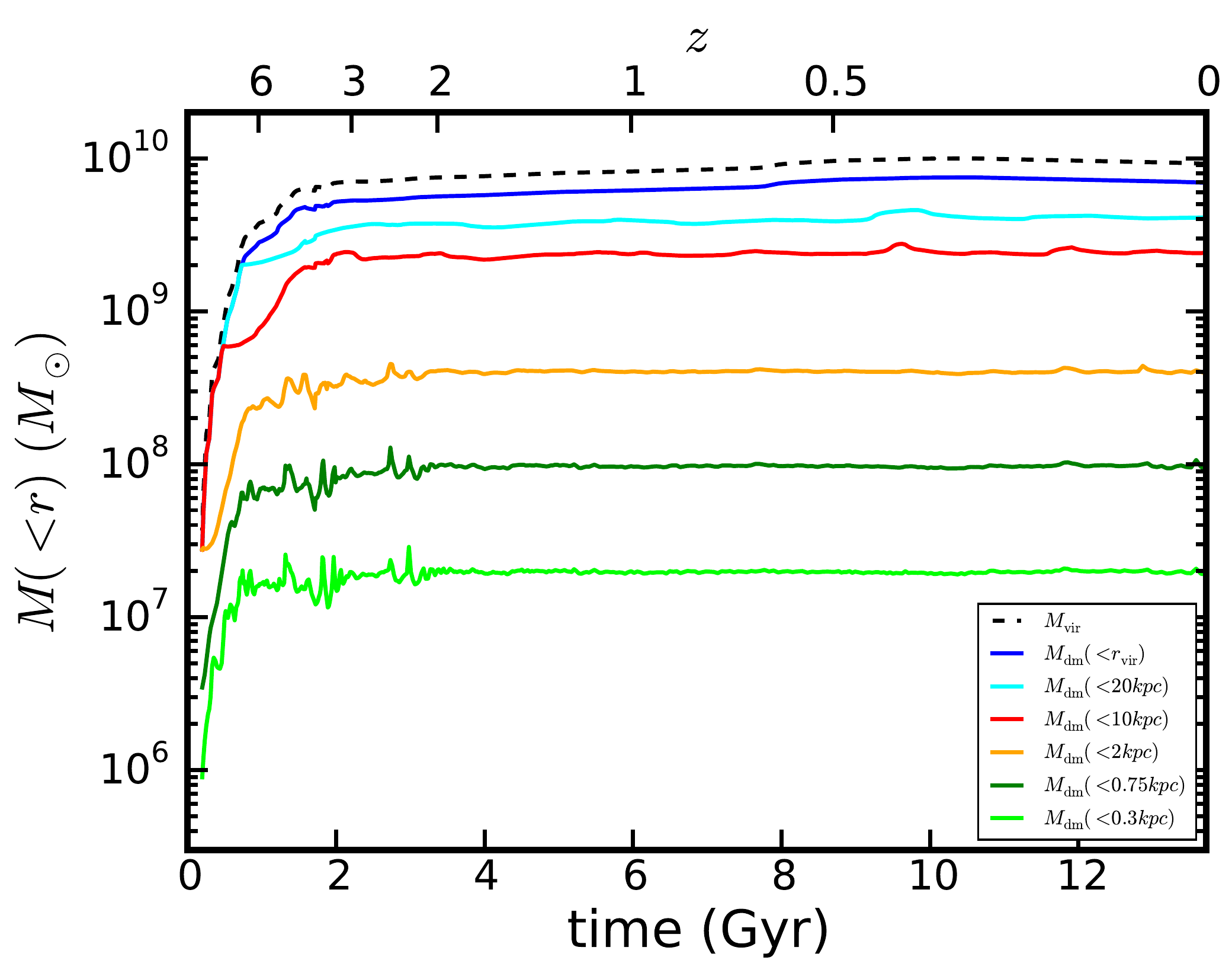}
 \caption{The evolution of the dark matter of the Dwarf collisionless run. Each line shows the dark matter mass contained inside a fixed physical radius.}
 \label{fig:massdmrun}
\end{figure}

Figure~\ref{fig:massdmrun} shows the evolution of the dark matter mass profile
for the Dwarf collisionless simulation. Each line shows the amount of dark
matter mass contained inside a fixed physical radius. At high redshift the halo shows a
characteristic fast halo mass increase followed by a very shallow evolution at
high redshift. Notice how the inner parts of the profile takes a bit more time
to settle down. Below redshift $z\sim2.5$ the inner part of the halo does not show
any significant perturbation as there is no significant accretion or merger
\citep{Diemand:2007, Diemer:2013}. 

\section{Convergence}
\label{app:convergence}
In this Section we present a convergence study that we have performed for the Dwarf galaxy halo. We have run a lower resolution version of all the Dwarf hydrodynamical runs discussed above. The only difference between the high and low resolution runs are the different particle masses and softenings used in the runs. The star formation density threshold, $n_{\rm sf}$, is also slightly different between the runs, $10$ $cm^{-3}$ for low res and $100$ $cm^{-3}$ for high res. All other code and physical parameters are exactly the same as for the high resolution runs. In Table~\ref{tab:simsinfolr} all the relevant parameters of these lower resolution runs can be found.

\begin{figure*} 
  \begin{center}
  \includegraphics[width=0.45\textwidth]{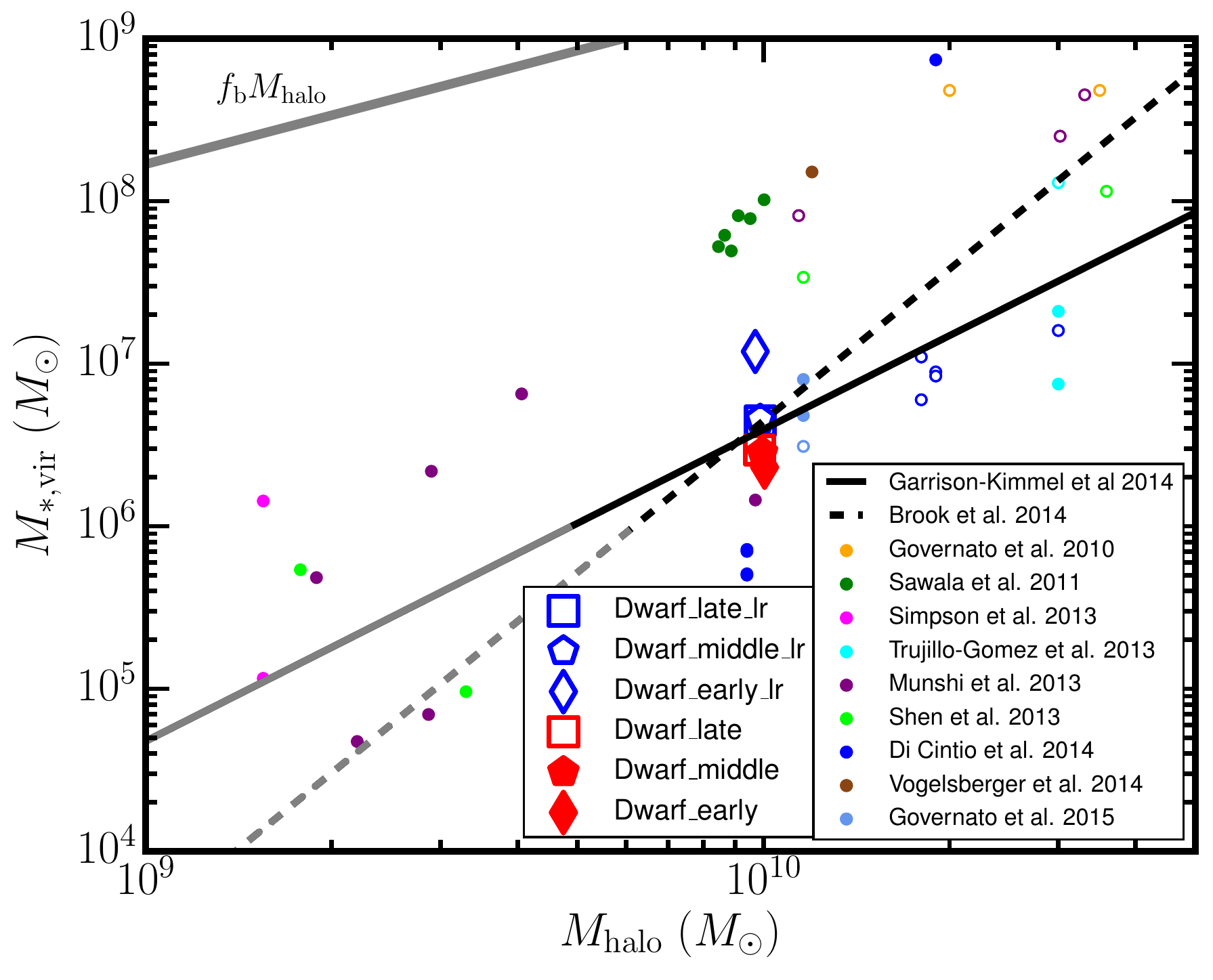}
  \includegraphics[width=0.45\textwidth]{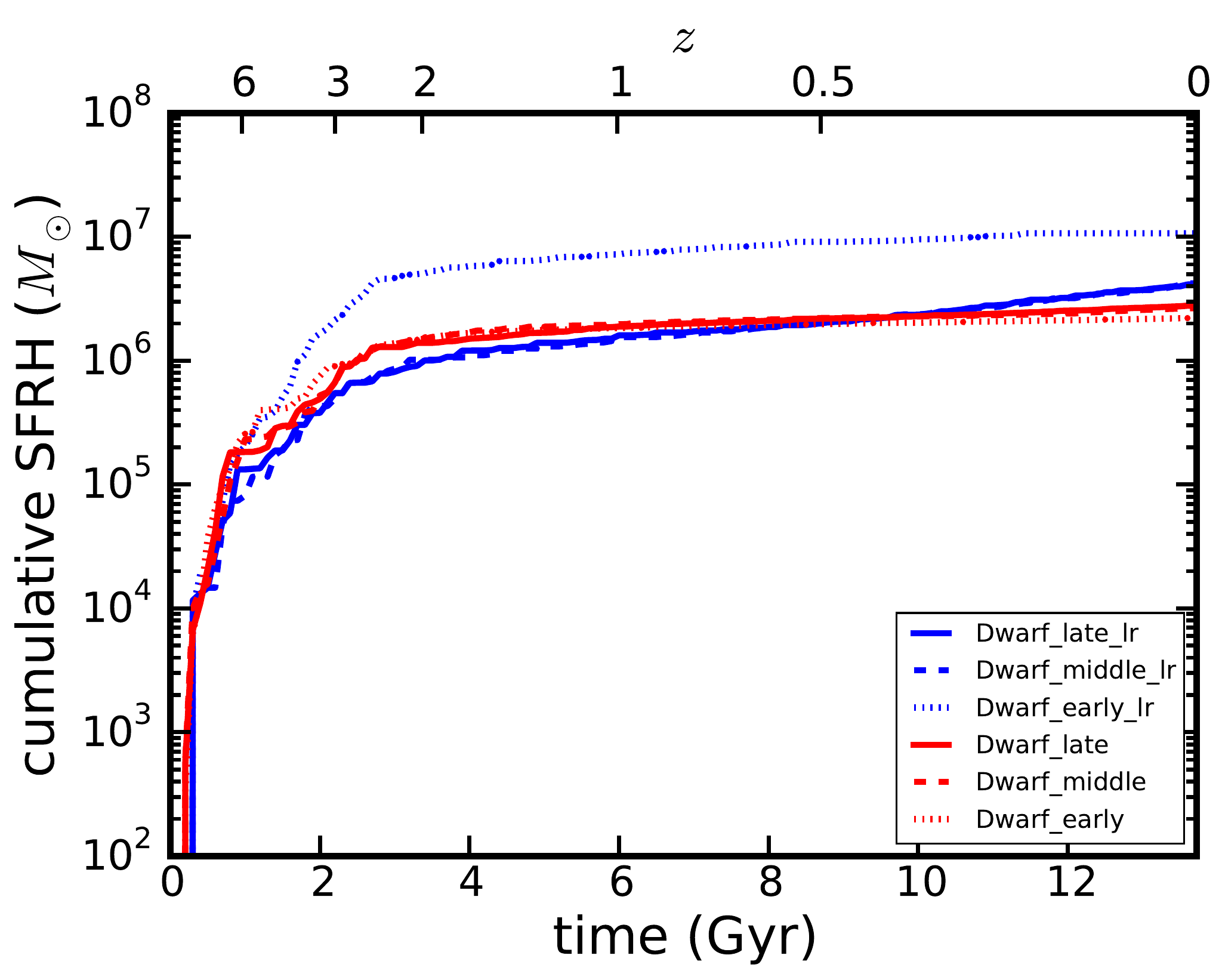}\\
  \includegraphics[width=0.45\textwidth]{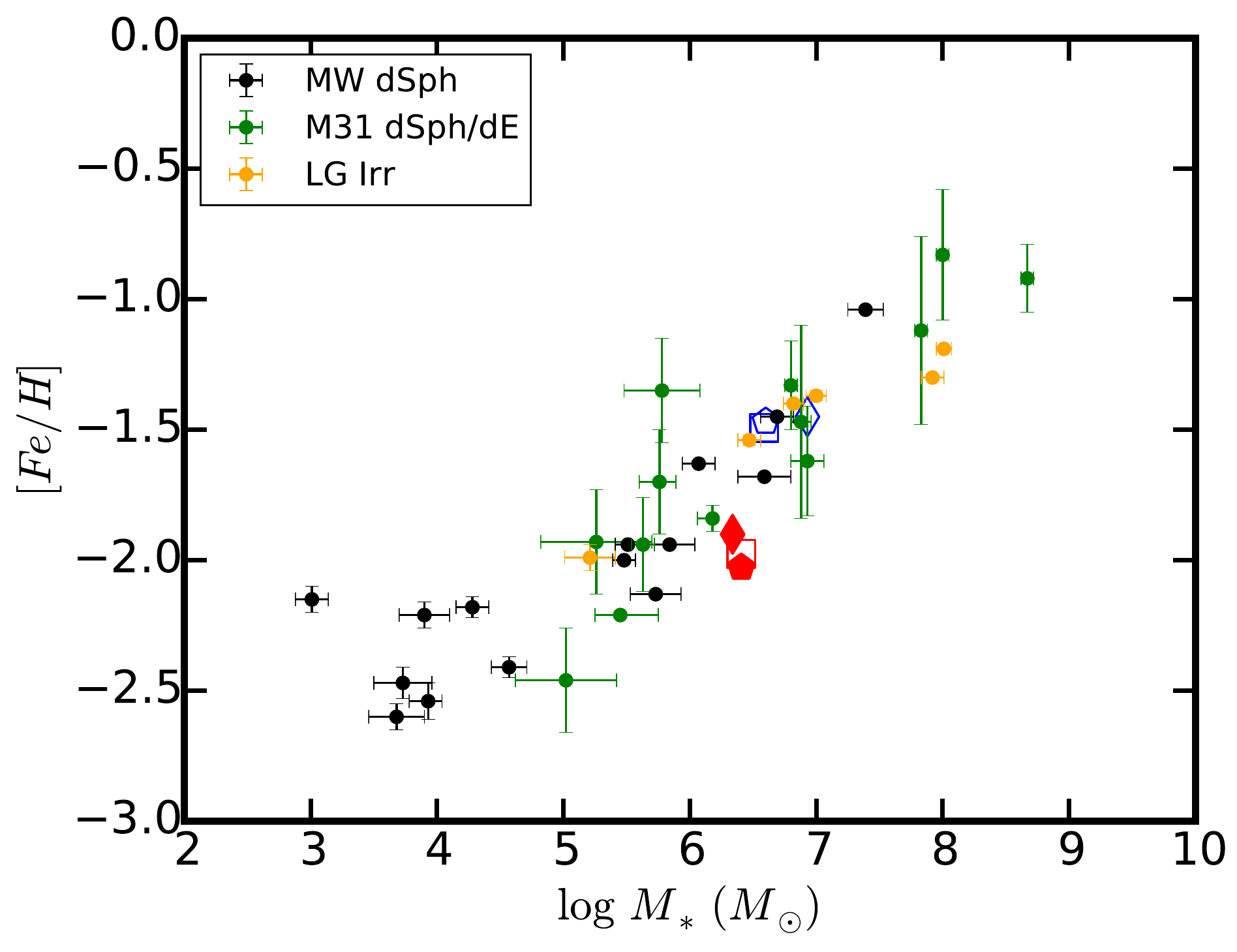}
  \includegraphics[width=0.45\textwidth]{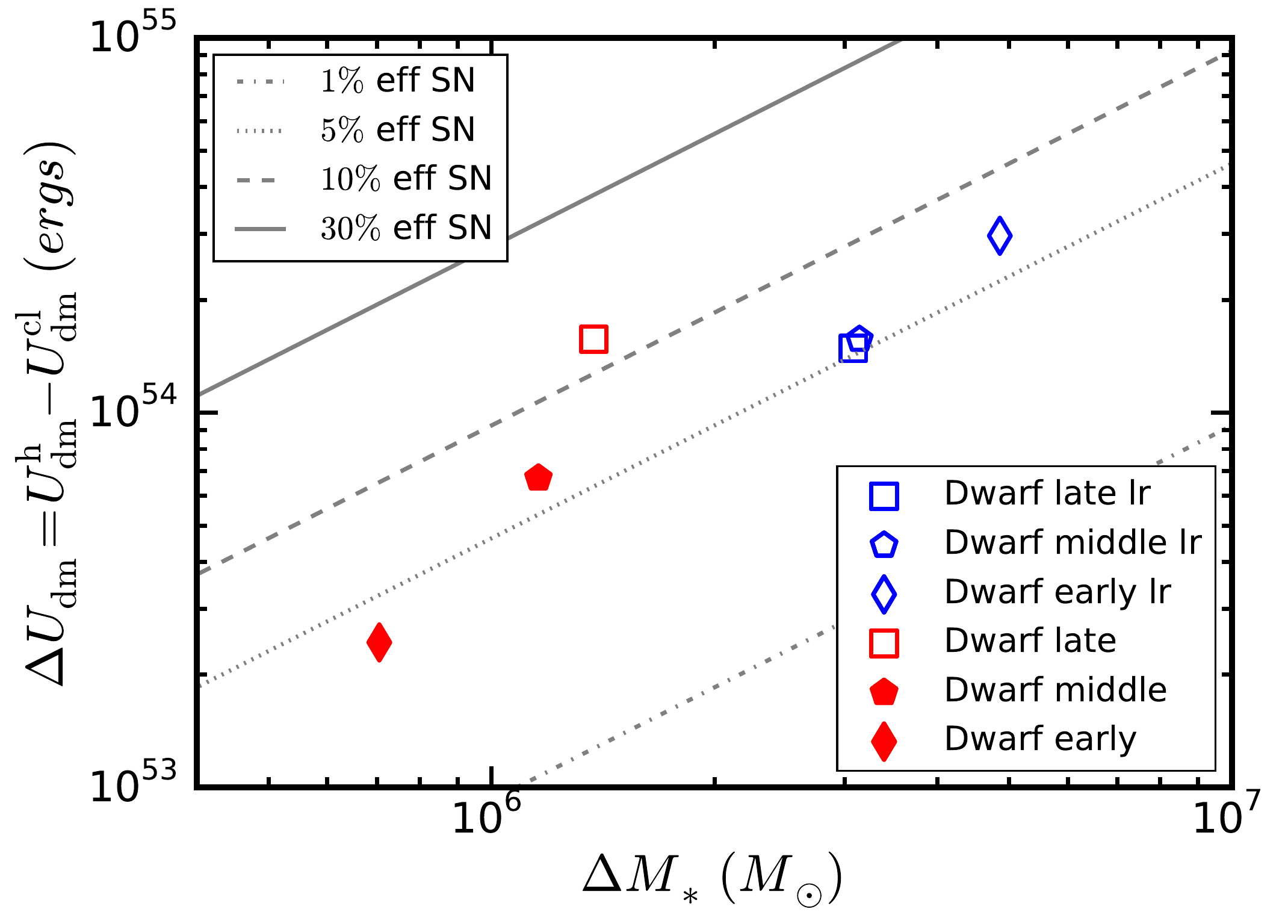}
  \end{center}
  \caption{Convergence tests. High (red) and low (blue) resolution simulations.
  Upper left: The stellar mass - halo mass relation. Upper right: Cumulative star
  formation history.  Lower left: Metallicity versus stellar mass.  Lower right: 
  Energy considerations in the formation of the dark matter cores in a $10^{10} \msun$
  halo. See text for details.}
  \label{fig:restest}
\end{figure*}

The different panels of Figure~\ref{fig:restest} illustrate the differences between the runs. The main difference that we found is that the low resolution runs have slightly higher stellar masses (upper left panel of Figure~\ref{fig:restest}). This can be understood by looking at the SFR histories (upper right panel in Figure~\ref{fig:restest}, blue lines stand for low resolution runs and red lines for the high resolution ones). The main difference observed between resolutions is the steeper slope of the cumulative star formation history at lower redshift. This produces higher stellar masses at $z=0$ for the lower resolution runs. 
We think that this is because the minimum amount of star formation that is possible is set by the gas particle resolution. Therefore the minimum amount of star formation is higher in the lower resolution runs. It is remarkable that all galaxy trends with size and metallicity hold regarding of resolution, so the galaxies seems just move along these relations (lower left panel of Figure~\ref{fig:restest}). We have also re-run our dwarf galaxy low resolution initial conditions using exactly the same code to check for pure stochastic differences. The scatter found in all the properties studied in this paper was similar to the one that we found when we change the feedback implementation and/or the softening values. These authors suspect that these differences will decrease at higher resolution, though higher resolution runs will certainly be required in order to test this conjecture.

Finally, concerning the core formation and energy considerations, low resolution runs also form a core which seems to be directly connected with its star formation rate at low redshifts ($z\lesssim2$). In general these cores are more prominent than their high resolution counterparts due to the higher star formation rate discussed above. Remarkably when we plot the energy requirements to form these cores versus the amount of energy obtained from supernova feedback below $z=2$ (lower right panel of Figure~\ref{fig:restest}), they lie in the same range of efficiencies as their high resolution counterparts.

Although it is not possible to claim full convergence for our high resolution runs from these results, we think that they are at least quite encouraging and definitely an improvement from other approaches in which parameters of the sub-grid physics must be tuned at each resolution. 

\begin{table*}
\centering
\begin{footnotesize}
\begin{tabular}{l|cccc}  
  \hline
  Parameter & Dwarf\_dm\_lr  & Dwarf\_late\_lr & Dwarf\_middle\_lr & Dwarf\_early\_lr\\
    & (Collisionless)  & (Hydro: Feed-M) & (Hydro: Feed-M-soft) & (Hydro: Feed-V) \\
  \hline
  1) $m_{\rm p}^{\rm dm}$ ($M_{\sun}$) & $1.21\times10^{4}$ & $1.01\times10^{4}$& $1.01\times10^{4}$& $1.01\times10^{4}$\\
  2) $\epsilon_{\rm dm}$ ($pc$) & $35$ & $35$ & $35$ & $35$ \\
  3) $m_{\rm p}^{\rm bar}$ ($M_{\sun}$) & -- & $2.04\times10^{3}$ & $2.04\times10^{3}$& $2.04\times10^{3}$\\
  4) $\epsilon_{\rm gas}^{\rm min}$ ($pc$) & -- & $2.0$ & $35$ & $2.0$\\
  \hline
  5) $\mvir$ $(\msun)$ & $9.48\times 10^{9}$ & $7.60\times 10^{9}$& $7.60\times 10^{9}$ & $7.46\times 10^{9}$ \\
  6) $\vmax$ $(km/s)$ & $37.31$ & $32.79$& $32.79$ & $33.56$ \\
  7) $\rvir$ $(kpc)$ & $54.99$ & $51.08$ & $51.08$ & $50.77$\\
  8) $\fbar$ & -- & $0.0166$ & $0.0171$ & $0.0137$\\
  9) $\fgas$ & -- & $0.0160$ & $0.0165$ & $0.0121$ \\
  10) $\fstar$ & -- & $0.0006$ & $0.0006$ & $0.0016$\\
  11) $\mobstar$ $(\msun)$ & -- & $4.1\times 10^{6}$ & $4.2\times 10^{6}$ & $1.0\times 10^{7}$\\
  12) $\reobstar$ $({\rm kpc})$  & -- & $0.783$& $0.881$ & $1.311$ \\
  13) $\FeH$  & -- & $-1.493$ & $-1.468$ & $-1.450$\\
  14) $\mtotTT$ $(\msun)$  & $3.212\times 10^{7}$ & $1.285\times 10^{7}$ & $1.014\times 10^{7}$ & $4.088\times 10^{6}$\\
  15) $\mdmTT$ $(\msun)$   & $2.414\times 10^{7}$ & $7.366\times 10^{6}$ & $7.132\times 10^{6}$ & $3.703\times 10^{6}$\\
  16) $\mbarTT$ $(\msun)$  & -- & $5.483\times 10^{6}$ & $3.002\times 10^{6}$ & $3.861\times 10^{5}$\\
  17) $\mgasTT$ $(\msun)$  & -- & $5.083\times 10^{6}$ & $2.635\times 10^{6}$ & $0.0$\\
  18) $\mstarTT$ $(\msun)$ & -- & $4.004\times 10^{5}$ & $3.677\times 10^{5}$ & $3.861\times 10^{5}$\\
   \hline
\end{tabular}
\end{footnotesize}
\caption{Simulations data for the low resolution convergence tests. First column stand for the different parameters studied for each simulation. In Columns 2-9 results for the simulations presented in this work are shown. 
Row 1: dark matter particle mass in the high resolution region in solar masses. 
Row 2: fixed gravitational softening used for the dark matter particles in physical parsecs. 
Row 3: baryon particle mass in the high resolution region in solar masses.
Row 4: minimum baryonic force softening in parsecs (minimum SPH smoothing lengths are comparable or smaller). Recall, force softenings are adaptive (mass resolution is fixed).
Row 5: virial mass in solar masses defined at the overdensity at which the spherical top hat model predicts virialization \citep{Bryan:1998}.
Row 6: maximum circular velocity in $km/s$.
Row 7: virial radius in kiloparsecs.
Row 8: virial baryon fraction, i.e., baryon mass inside the virial radius over the virial mass.
Row 9: virial gas fraction, i.e., gas mass inside the virial radius over the virial mass.
Row 10: virial stellar fraction, i.e., stellar mass inside the virial radius over the virial mass.
Row 11: stellar mass in solar masses. This is the stellar mass of the central galaxy.
Row 12: effective stellar mass radius, i.e., half stellar mass radius in kiloparsecs.
Row 13: stellar iron over hydrogen ratio. Mass weighted iron over hydrogen ratio for the dwarf stellar mass component. 
Row 14: total mass inside $500$ parsec in solar masses.
Row 15: dark matter mass inside $500$ parsec in solar masses.
Row 16: baryon mass inside $500$ parsec in solar masses.
Row 17: gas mass inside $500$ parsec in solar masses.
Row 18: stellar mass inside $500$ parsec in solar masses.}
\label{tab:simsinfolr}
\end{table*}
\end{document}